\documentclass{aa}
\usepackage[utf8]{inputenc}
\usepackage{newunicodechar}
\newunicodechar{₂}{$_2$}
\usepackage{booktabs}
\usepackage{algorithm2e}
\usepackage{pdflscape}
\usepackage{array}
\usepackage{gensymb} 
\usepackage{amsmath}
\usepackage{adjustbox}
\usepackage{lipsum} 
\usepackage{rotating}
\usepackage{float} 
\usepackage{graphicx}
\usepackage[english]{babel}
\usepackage[pagebackref]{hyperref}
\hypersetup{colorlinks=true, citecolor=blue,linkcolor=black,urlcolor=blue}
\usepackage{natbib}
\usepackage{subfig}
\usepackage{txfonts}
\usepackage{xcolor} 
\usepackage{appendix}

\begin{document}

   \title{The role of young and evolved stars in the heating of dust in local galaxies}

   \subtitle{}

\author{
    Vidhi Tailor\inst{1,2}
    \and
    Viviana Casasola\inst{1}
    \and
    Francesca Pozzi\inst{2,3}
    \and
    Francesco Calura\inst{3}
    \and
    Simone Bianchi\inst{4}
    \and
    Monica Relano\inst{5,6}
    \and
    Jacopo Fritz\inst{7,1}
   \and
   Frédéric Galliano\inst{8}
    \and
   Matteo Bonato\inst{1}
    \and
    Maritza A. Lara-López\inst{9,10}
    \and
    Evangelos Dimitrios Paspaliaris\inst{4}
    \and
    Alberto Traina\inst{3}
}
\institute{
    INAF - Istituto di Radioastronomia, Via Gobetti 101, I-40129, Bologna, Italy \\ \email{vidhiritesh.tailor@unibo.it}
    \and
    Dipartimento di Fisica e Astronomia, Alma Mater Studiorum Università di Bologna, Via Piero Gobetti 93/2, I-40129, Bologna, Italy
    \and
    INAF – Osservatorio di Astrofisica e Scienza dello Spazio di Bologna, Via Gobetti 93/3, 40129, Bologna, Italy
    \and
   INAF - Osservatorio Astrofisico di Arcetri, Largo E. Fermi 5, 50125, Florence, Italy
    \and
   Dept. Física Teórica y del Cosmos, Universidad de Granada, Spain
   \and
   Instituto Universitario Carlos I de Física Te\'{o}rica y Computacional, Universidad de Granada, 18071, Granada, Spain
   \and
    Instituto de Radioastronomía y Astrofísica (IRyA), Universidad Nacional Autónoma de México (UNAM), Antigua Carretera a Pátzcuaro, 8701, Ex-Hda. San José de la Huerta, Morelia, Michoacán, 58089, México
    \and
    Université Paris-Saclay, Université Paris Cité, CEA, CNRS, AIM, 91191, Gif-sur-Yvette, France
     \and
    Departamento de Física de la Tierra y Astrofísica, Fac. de C.C. Físicas, Universidad Complutense de Madrid, E-28040 Madrid, Spain
    \and
   Instituto de Física de Partículas y del Cosmos, IPARCOS, Fac. C.C. Físicas, Universidad Complutense de Madrid, E-28040 Madrid, Spain
}

   \date{\today}

\abstract
  % context heading 
  {Dust is a fundamental component of the interstellar medium (ISM) and plays a critical role in shaping galaxy evolution. Dust grains influence the ISM by cooling the gas, altering its chemistry, and absorbing stellar radiation, re-emitting it at longer wavelengths in the FIR and sub-mm regimes. The cold dust component, which constitutes the majority of the dust mass, is primarily heated by stellar radiation, with contributions from both young, massive stars and the diffuse emission from older stellar populations. It is essential to discern how dust is heated in order to better understand the relationship between stellar populations and their surrounding environments.}
  % aims heading 
  {This study aims to identify the dominant heating mechanisms responsible for the cold dust component in typical nearby spiral galaxies and to explore the contributions of both young and evolved stellar populations to dust heating.}
  % methods heading 
  {Using a sample of 18 large, face-on spiral galaxies from the DustPedia project, we employ two complementary approaches. 
   In the first method, we study the correlations between dust temperature (\(T_{\rm dust}\)), star-formation rate (SFR) surface density (\(\Sigma_{\rm SFR}\)), and stellar mass surface density (\(\Sigma_{M_*}\)). 
   In the second method, we explore the relationship between \(T_{\rm dust}\) and dust mass surface density (\(\Sigma_{\rm dust}\)).}
  % results heading 
  {By analyzing the median temperature radial profile, we find that \(T_{\rm dust}\) peaks at $\sim$24~K at the galaxy center, decreasing to $\sim$15~K toward the galaxy outskirts. 
  Our analysis shows similar \(T_{\rm dust}\) in galaxies with and without a central active galactic nucleus (AGN), suggesting that AGN activity does not significantly influence \(T_{\rm dust}\) values and distribution on the spatial scales covered by our data, ranging from 0.3 to 3~kpc. 
  For $\sim$72$\%$ of the galaxies in our sample, the methods consistently identify the primary dust heating source. However, when considering the entire galaxy sample, our analysis suggests that there is no single dominant heating mechanism; both young and evolved stars contribute to dust heating, with their relative contributions varying across galaxies. }
  % conclusions heading (optional), leave it empty if necessary 
  {}


   \maketitle
\section{Introduction}

Interstellar dust, though comprising only $\sim$1$\%$ of the ISM by mass \citep[e.g.,][]{Draine_2011}, has a major impact on galaxy evolution by regulating heating, cooling, and chemistry. Dust grains are primarily produced in the ejecta of evolved stars, such as AGB stars and supernovae, and are injected into the ISM via stellar winds and explosive events \citep[e.g.,][]{Gehrz_1989,Matsuura_2009,Nanni_2013,Nanni_2014,Dell'Agli_2017,Ventura_2018}. Once injected into the ISM, dust grains grow by accreting metals and coagulating into larger particles, but can also be fragmented or destroyed by supernova-driven shocks \citep[e.g.,][]{Jones_1996,Hirashita_2019}.
Over time, energetic gas particles erode dust through sputtering, returning the material to the gas phase.
These processes of formation, growth, and destruction highlight the role dust plays in the ISM cycle
\citep[e.g.,][]{Calura_2008,Asano_2013,Gallino_2021}.

Dust grains also affect the spectral energy distribution (SED) of galaxies by absorbing ultraviolet (UV)/optical light, then re-emitting it at longer wavelengths, in the infrared (IR) and submillimeter (sub-mm) range.
The degree to which these processes affect the ISM is highly dependent on both the size and composition of the dust grains \citep[e.g.,][]{Jones_2017,Bianchi_2019,Galliano_2022}. 
Moreover, dust influences a galaxy's metallicity, as a significant fraction of the metals produced through stellar nucleosynthesis is returned to the ISM via stellar mass loss or through SNe, condensing these metals into dust grains. As dust grains are composed of metals, their evolution is intrinsically linked to metallicity \citep[e.g.,][]{Hu_2023}. 
Dust grains also play a pivotal role in catalysing the formation of molecular hydrogen (H$_2$), the most abundant molecule in ISM.
Dust protects H$_2$ molecules from UV radiation that would otherwise cause their dissociation \citep[e.g.,][]{Hollenbach_1971,Dwek_1987}.
In fact, molecular shielding is crucial in the SF process, as it allows the gas to cool down, leading to the development of high-density regions that are decisive for triggering the onset of SF \citep[e.g.,][]{Kennicutt_2007,Bigiel_2011}.  

Dust grains in galaxies exhibit a range of temperatures, influenced by their size and the intensity of the radiation field to which they are exposed \citep{Li_Dreaine_2001}. 
Grains in thermal equilibrium can be heated by the diffuse radiation originating from both evolved stellar populations and young stars. However, when embedded in star-forming regions, they are primarily heated by intense radiation from newly formed UV-emitting stars, closely linked to recent SF \citep[e.g.,][]{Nersesian_2019}.
Small grains experience stochastic heating and can reach very high temperatures (up to 1000 K or more) due to single-photon absorption events, with their temperature depending on their size, composition, and the energy of the absorbed photon.
In general, large grains, which dominate the dust mass in galaxies, reach relatively stable equilibrium temperatures of $\sim$15--30~K and predominantly emit in the far infrared (FIR), where they contribute to the majority of observed dust emission \citep[e.g.,][]{Orellana_2017, Pozzi_2021}.

The \textit{Herschel Space Observatory} \citep[][]{Pilbratt_2010}, together with other space-based IR telescopes 
(e.g., \textit{Spitzer}, \citeauthor{Spitzer}~\citeyear{Spitzer}; WISE, \citeauthor{WISE}~\citeyear{WISE}; JWST \citeauthor{JWST}~\citeyear{JWST})
has revolutionised our understanding of the ISM properties and, in particular, of the dust. \textit{Herschel} observations provided maps of dust emission with a resolution  down to 6\arcsec - 36\arcsec, over a wide wavelength range (\(70 - 500\,\mu\)m), allowing detailed analyses of fluxes, masses, and temperatures of dust within galaxies. \textit{Herschel} observations have been crucial for investigating the cold-dust component \citep[$T$$\sim$15--20~K,][]{Galliano_2018}, which does not emit significantly at shorter wavelengths \citep[$<$~50~$\mu$m,][]{Draine_2003}, making FIR observations indispensable for nearby galaxies \citep[e.g.,][]{Ruyer2015,Devis_2019,Ner_2021}.

Thanks to \textit{Herschel}, several efforts have been devoted to characterizing dust in nearby galaxies. For example, \cite{Boselli_2012} studied the relationships between different \textit{Herschel} FIR color indices and their connection with various physical properties of the galaxies. They concluded that the emission of the cold dust is regulated by the properties of the interstellar radiation field (ISRF), in addition to the gas-phase metallicity. \cite{Bendo_2015} analyzed the variations in the 160/250~$\mu$m and 250/350~$\mu$m surface brightness ratios observed in a sample of nearby galaxies and compared them with tracers of SF and older stellar populations, identifying a broad variation in the heating of the dust observed in the 160--350~$\mu$m range. 
For some galaxies, they found evidence that emission at $\leq$~160~$\mu$m originates from dust heated by star-forming regions, for others, the emission at $\geq$ 250~$\mu$m originates from dust heated by the older stellar population, and for others both stellar populations may contribute equally to the global dust heating or the results are inconclusive \citep[see also, i.e.,][]{Smith_2010,Hughes_2014}. \cite{Smith_2012} analyzed \textit{Herschel} observations of M31 and found that dust in the bulge is primarily heated by evolved stars, with a weak correlation between dust temperature and SF in the outer disk. 
Similarly, \cite{Viaene_2014} examined dust scaling relations in M31, highlighting how dust content is regulated by stellar mass and SF history. More recently, \citet{Nersesian_2019} and \citet{Bonato_2024} found that dust heating in early-type galaxies is mainly due to old stars, whereas young stars progressively contribute more for spiral galaxies, although the contribution of old stars remains important, if not dominant.

Beyond empirical methods, radiative transfer (RT) models are essential for studying dust heating by simulating the propagation of photons through the ISM. These models account for complex geometries and varying optical depths, making them valuable for understanding radiation-dust interactions.
For example, RT models like SKIRT \citep{Baes_2011} utilize 3D Monte Carlo simulations to model absorption, scattering, and thermal emission of dust grains, allowing detailed studies of dust heating by different stellar populations \citep[e.g.,][]{DeLooze_2014,Viaene_2017,Nersesian_2020a,Nersesian_2020b,Verstocken_2020}. Although complete RT simulations offer high precision, they can be computationally expensive and time-consuming. 

In this context, the present study aims to understand the dominant heating source of the cold dust component in typical star-forming galaxies. 
We consider a subset of 18 large, nearby, face-on spiral galaxies from the DustPedia sample \citep{Davies_2017}, as presented in \citeauthor{Casasola_2017}~(\citeyear{Casasola_2017}, hereafter C17). We analyze the dust properties (temperature and mass) and spatially resolved dust correlations, in a range of physical scales between 0.3 and 3.4~kpc.
The correlations studied are between dust temperature ($T_{\rm {dust}}$), dust mass surface density ($\Sigma_{\rm {dust}}$), star-formation rate (SFR) surface density \((\Sigma_{\text{SFR}})\), and stellar mass surface density ($\Sigma_{\rm M_*}$). Analyzing these dust properties and correlations, we explore
how different stellar populations and galaxy properties contribute to dust heating.

The structure of the paper is as follows. 
Sect.~\ref{sec:sample} describes the galaxy sample and the dataset used in this study.
Sect.~\ref{sec:results} presents the $T_{\rm {dust}}$ radial profiles derived for each sample galaxy and the average $T_{\rm {dust}}$ radial profile for the entire sample, with a focus on the contribution of the AGN to the heating of dust and analyse the dust correlations. Sect.~\ref{sec:discussion} discusses the results, and Sect.~\ref{sec:conclusions} summarizes the study and proposes future perspectives.

\begin{table*}[ht!]
    \caption{Main properties of the sample galaxies.}
    \centering
    \begin{tabular}{lcccccccl}
        \hline\\      
        Galaxy &  \(\alpha_{\rm J2000}\) &  \(\delta_{\rm J2000}\) & RC3 type & \(D_{25}\) &  Distance &  \textit{i} &  Nuclear Activity \\
        &  [\(^{\rm h}\) \(^{\rm m}\) \(^{\rm s}\)] &  [\(^{\circ}\) \(^{\prime}\) \(^{\prime\prime}\)] & &  [\(^{\prime}\)] & [Mpc] &  [\(^{\circ}\)] &   \\
        (1) & (2) & (3) & (4) & (5) & (6) & (7) & (8)\\
        \hline
        NGC 300         & 00 54 53.4 & -37 41 03 & SA(s)d        & 19.5 & 1.99            & 43.0  & --$^{(1)}$ \\
        NGC 2403        & 07 36 51.1 & +65 36 03 & SAB(s)cd      & 20.0 & 3.18            & 62.9  &H{\sc ii}\textsuperscript{(d)} \\
        IC 342          & 03 46 48.5 & +68 05 47 & SAB(rs)cd     & 20.0 & 3.39             & 31.0  & H{\sc ii}\textsuperscript{(d)} \\
        NGC 7793        & 23 57 49.7 & -32 35 28 & SA(s)d        & 10.5 & 3.60             & 49.6  & H{\sc ii}\textsuperscript{(e)} \\
        NGC 3031 (M 81) & 09 55 33.1 & +69 03 44 & SA(s)ab       & 21.4 & 3.61             & 59.0  & LINER\textsuperscript{(d)}\\
        NGC 6946        & 20 34 52.2 & +60 09 14 & SAB(rs)cd     & 11.5 & 4.51\textsuperscript{(a)} & 32.6  & H{\sc ii}\textsuperscript{(c)}\\
        NGC 4736 (M 94) & 12 50 53.0 & +41 07 13 & (R)SA(r)ab    & 7.8  & 4.59             & 41.4  & LINER\textsuperscript{(d)}\\
        NGC 5236 (M 83) & 13 37 00.9 & -29 51 57 & SAB(s)c       & 13.5 & 4.66            & 24.0  &  H{\sc ii} / LINER\textsuperscript{(e)} \\
        NGC 3621        & 11 18 16.5 & -32 48 51 & SA(s)d        & 9.8  & 6.70             & 64.7  & LINER / Seyfert\textsuperscript{(e)} \\
        NGC 5457 (M 101)& 14 03 12.6 & +54 20 57 & SAB(rs)cd     & 24.0 & 6.95           & 18.0  & H{\sc ii}\textsuperscript{(d)} \\
        NGC 5194 (M 51) & 13 29 52.7 & +47 11 43 & SA(s)bc pec   & 13.8 & 7.55\textsuperscript{(b)} & 42.0  & Seyfert\textsuperscript{(d)} \\
        NGC 925         & 02 27 16.5 & +33 34 44 & SAB(s)d       & 10.7 & 8.67             & 66.0  & H{\sc ii}\textsuperscript{(d)} \\
        NGC 628 (M 74)  & 01 36 41.8 & +15 47 00 & SA(s)c        & 10.0 & 8.83\textsuperscript{(b)} & 7.0   & H{\sc ii}\textsuperscript{(c)} \\
        NGC 5055 (M 63) & 13 15 49.2 & +42 01 45 & SA(rs)bc      & 11.8 & 8.99            & 59.0  &  LINER\textsuperscript{(d)}\\
        NGC 4725        & 12 50 26.6 & +25 30 03 & SAB(r)ab pec  & 9.8  & 12.88           & 54.0  & Seyfert\textsuperscript{(d)} \\
        NGC 3521        & 11 05 48.6 & -00 02 09 & SAB(rs)bc     & 8.3  & 13.24            & 72.7  & LINER\textsuperscript{(c)} \\
        NGC 1097        & 02 46 19.0 & -30 16 30 & SB(s)b        & 10.5 & 15.78            & 46.0  & LINER\textsuperscript{(f)} \\
        NGC 1365        & 03 33 36.4 & -36 08 25 & SB(s)b        & 12.0 & 16.98           & 40.0  & Seyfert\textsuperscript{(f)}\\
        \hline
    \end{tabular}
    \label{tab:sample}
    \vspace{2mm}
    \tablefoot{Main properties of the sample galaxies. The columns list the following: (1) galaxy name; (2) right ascension (\(\alpha_{\rm J2000}\)) and (3) declination (\(\delta_{\rm J2000}\)) in J2000 coordinates; (4) RC3 classification type; (5) optical diameter; (6) galaxy distance; (7) galaxy inclination angle; (8) nuclear activity. The galaxies are arranged in order of increasing distance. Distances are primarily taken from the Cosmicflows-3 database \citep{Tully_2016}, except where noted: \textsuperscript{(a)}~\cite{Dist_N6946}; \textsuperscript{(b)}~\cite{Dist_N5194_N628}. The nuclear activity classification is from: \textsuperscript{(c)}~\cite{Goulding_Class_c};\textsuperscript{(d)}~\cite{Lem_cass_d};\textsuperscript{(e)}~\cite{Menzes_class_e}; \textsuperscript{(f)}~\cite{Thomas_class_f}. The nuclear activity is classified as H{\sc ii}, Seyfert, and LINER. Two galaxies (NGC~5236, NGC~3621) have ambiguous nuclear classifications. $^{(1)}$ Classification not available. For source references of other galaxy properties, see C17.}   
\end{table*}

\section{Sample selection and dataset}
\label{sec:sample}
The DustPedia project contains 875 apparently large ($D_{25}$\footnote{$D_{25}$ is the major axis isophote at which the optical surface brightness falls beneath 25~mag~arcsec$^{-2}$ (this is the diameter of the galaxy if it is a disk). We also use $R_{25} = D_{25}/2$.}~$>1^\prime$) and nearby \((\leq 40 \text{ Mpc})\) galaxies, all observed by \textit{Herschel}.
In addition to \textit{Herschel}, the DustPedia project provides a comprehensive database of multi-wavelength imagery and photometry (from UV to radio), surpassing similar surveys in both wavelength coverage and the number of galaxies observed. The data used in this work come from the DustPedia database\footnote{http://dustpedia.astro.noa.gr} and C17 \citep[see also,][for a detailed description of the DustPedia sample and database]{Davies_2017,Clark_2018}.

The galaxies sample of C17 consists of sources with a sub-mm minor-to-major axis ratio \((d/D)_{sub-mm}\) \(\geq 0.4\) and sub-mm diameter \(D_{sub-mm}\) \(\geq 9'\), which corresponds to approximately 15 resolution elements in the SPIRE--\( 500\; \mu m\) maps (FWHM = 36\arcsec).  These galaxies have a small (or moderate) inclination and have a diameter $D_{25}$~$\geq 7.8^{\prime}$. 
The sample spans Hubble types from Sa to Sd and includes 3 Sa–Sab, 5 Sb–Sbc, and 10 Sc–Sd galaxies, some of which exhibit structural features such as bars or signs of interaction.
Additionally, in our sample, 10 out of 18 galaxies are classified as low-luminosity active galactic nuclei (AGNs, $L_{\rm X} < 10^{42}$~erg~s$^{-1}$), including Seyferts and Low Ionization Nuclear Emission-line Regions (LINERs).
In contrast, the remaining ones have a H{\sc ii} nucleus.
However, for our analysis, we consider only Seyferts (4 galaxies) as AGN, since the energetic contribution from LINERs to dust heating is expected to be minimal (see Sect. \ref{sec:agn}).
The exception is NGC 300, for which no nuclear classification is available. Table~\ref{tab:sample} lists the main properties of the galaxy sample (for details on the sample, see C17).

In this study, we used $\Sigma_{\text{dust}}$, $\Sigma_{\text{SFR}}$, and $\Sigma_{\rm M_*}$ maps from C17.
As a novelty, we introduce $T_{\text{dust}}$ maps, not present in C17 but produced following the same methodology and assumptions used for $\Sigma_{\text{dust}}$.
All maps are corrected for inclination, $i$.

\subsection{Dust temperature and dust mass surface density}
\label{sec:mass-t}
Physical dust models assume specific compositions, densities, and shapes for dust grains, adopting heat capacities and optical properties from laboratory and/or theoretical studies \citep{Draine_li_2007, Jones_2013}. 
While earlier models often assumed spherical grains for simplicity, more recent models account for elongated grains \cite[e.g.,][]{Draine&Hensley_2023,Ysard_2024}. The temperature and emission of large grains in thermal equilibrium with a radiation field can be semi-analytically estimated, though numerical integration is required over realistic absorption efficiencies.
These physical models are calibrated by adjusting the grain size distribution and dust mass per H to simultaneously match observations of extinction, emission, and abundance, typically in a high-latitude Milky Way cirrus, where the underlying radiation field that is responsible for dust heating is known \citep[e.g.,][]{Mathis_1983,Bianchi_2017}.

Here, we recall the main steps used in the derivation of $\Sigma_{\rm dust}$ and $T_{\rm dust}$ maps of \citet{Casasola_2017}. The dust properties and grain size distributions were taken from the THEMIS\footnote{https://www.ias.u-psud.fr/themis/THEMIS$\_$model.html} (The Heterogeneous dust Evolution Model for Interstellar Solids) dust model \citep{Jones_2013,K_hler_2014,Jones_2017}, which, like other physical dust models, assumes that dust emission originates from grains exposed to a local ISRF, parameterized by the intensity $U$, where $U = 1$ corresponds to the Solar neighborhood ISRF. Within THEMIS, the dust grains consist of amorphous hydrocarbon (a-C(:H)) and amorphous olivine-type and pyroxene-type silicate with iron and iron sulfide nanoinclusions (a-Sil (Fe, FeS)), with silicates coated in a-C(:H) mantles \citep{Ysard_2015}. This dust model successfully reproduces the observed extinction, IR-to-millimeter thermal emission, and sub-mm opacity of Milky Way dust \citep[e.g.,][]{Fanciullo_2015,Ysard_2015}.
\begin{figure*}[h!]
    \centering
    \begin{minipage}{0.48\textwidth}
        \centering
        \includegraphics[width=\linewidth]{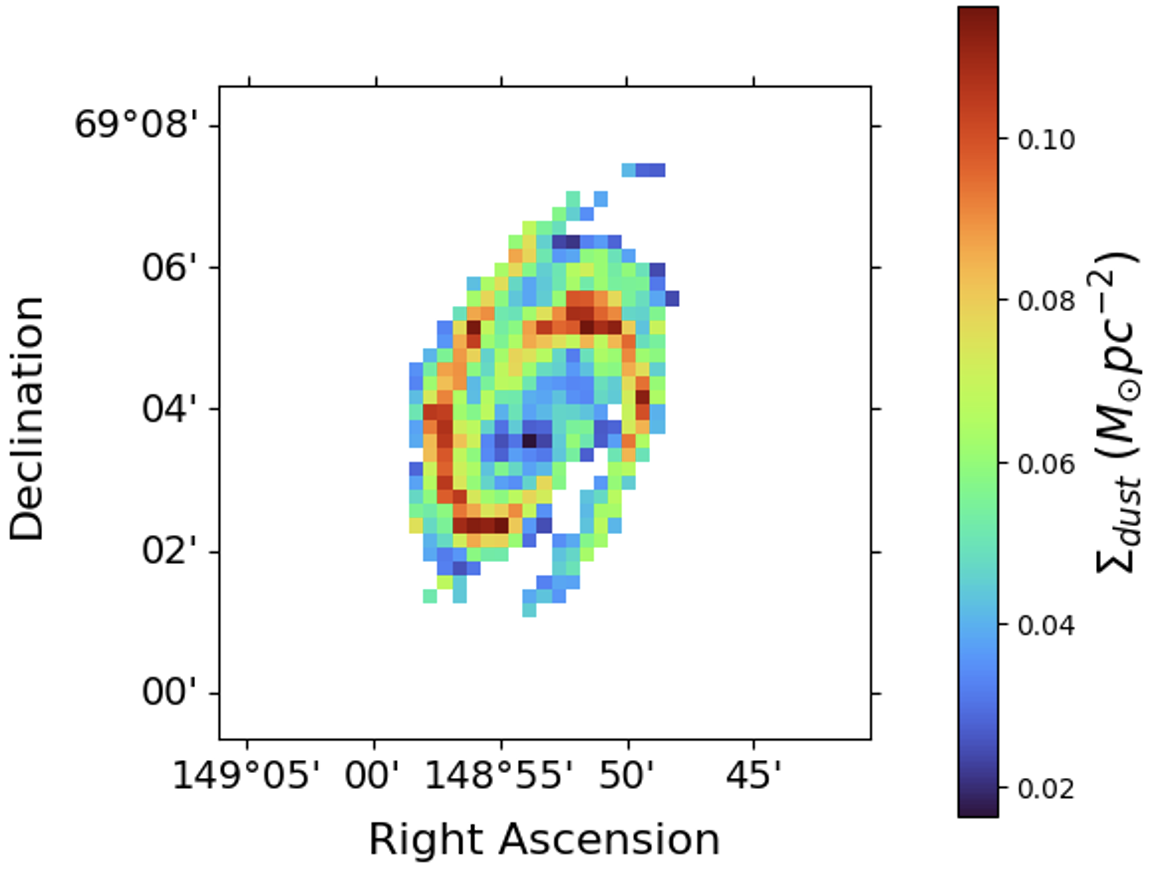}
    \end{minipage} 
    \hfill
    \begin{minipage}{0.48\textwidth}
        \centering
        \includegraphics[width=\linewidth]{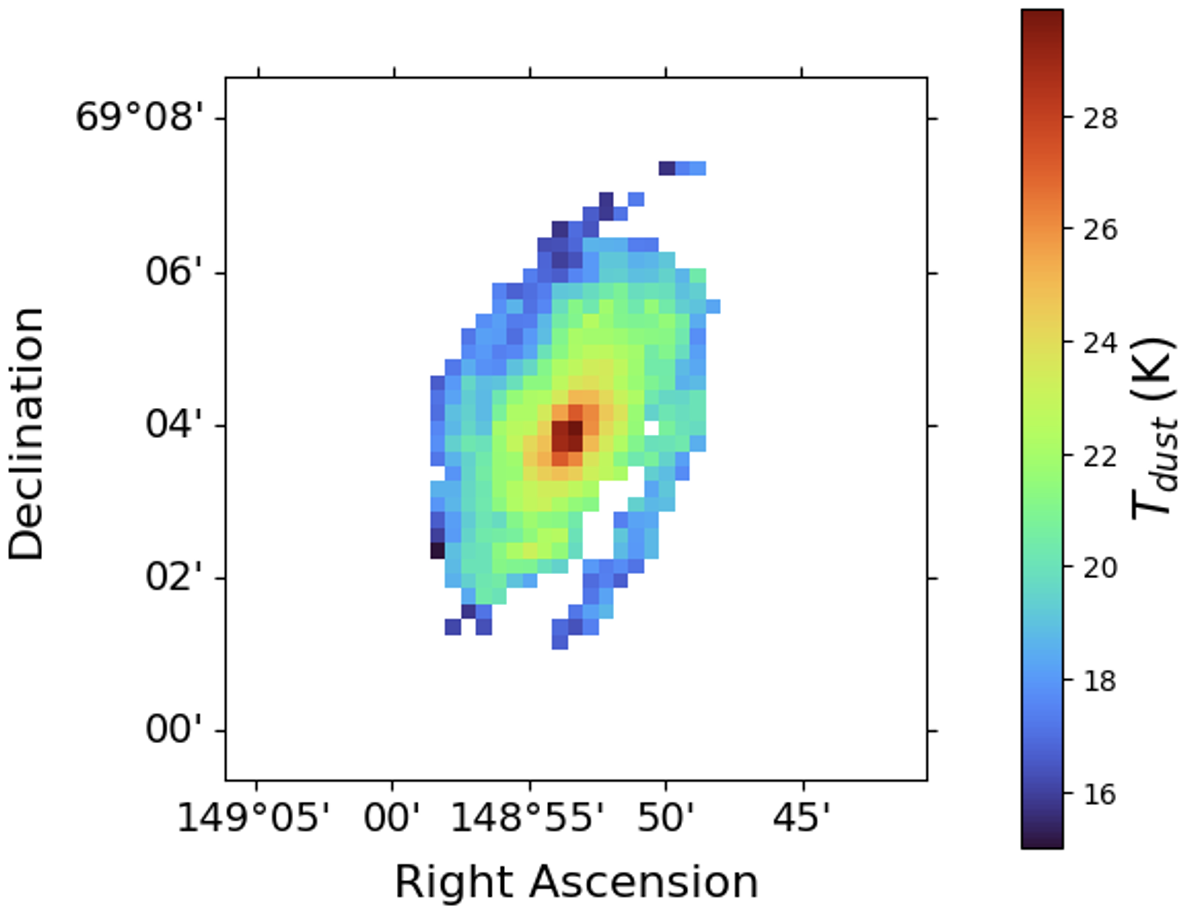}
    \end{minipage} 
    \caption{
    $\Sigma_{\text{dust}}$ (left) and $T_{\text{dust}}$ (right) maps of the sample galaxy NGC~3031 (M81) derived according to the prescriptions described in Sect.~\ref{sec:mass-t}. The maps are shown at the SPIRE-500 resolution (36\arcsec), corresponding to 0.6~kpc at the galaxy's distance. Only regions with signal-to-noise ratio (S/N) > 5 are shown.
    }
    \label{figdust}
\end{figure*}
 
The $\Sigma_{\text{dust}}$ and $U$ maps were obtained by fitting observed SEDs at each position within a galaxy, with the modeled SEDs convolved with \textit{Herschel} filter response functions.
Dust emission for various \(U\) values was computed using DustEM software \citep{Com_2011}. The radiation field $U$ was modeled as a single uniform ISRF, following a simplified version of the \cite{Draine_li_2007} approch,  assuming most of the dust mass is heated by a single diffused component (\(U = U_{min}\), in the notation of their method), with negligible contributions from higher-intensity components \citep[see][]{Dale_2012,Aniano_2012,Bianchi_2013}.  The spectral shape of the ISRF is fixed to that of the local Solar neighborhood \citep{Mathis_1983}, as is customary in SED fitting based on the \cite{Draine_li_2007} approach. Although this does not account for possible spectral hardening or reddening in different regions, such variations primarily affect the mid-IR regime and have limited impact on the FIR–sub-mm range used for our fits \citep[see also, ][]{Draine_2021,Bianchi_2022}.
SED fitting was performed using \textit{Herschel} bands with \(\lambda \geq 160 \, \mu m\) (PACS 160 $\mu$m, SPIRE 250, 350, 500 $\mu$m). When available, PACS \(100 \, \mu m\) data were included and PACS \(70 \, \mu m\) fluxes were treated as upper limits to exclude transiently heated grains \cite{R_my_Ruyer_2013}. For more information on the fitting process, refer to C17.
\(T_{\rm dust}\) is not directly derived from the SED fit, but calculated using the ISRF strength, parameterized by \(U\) \citep{Aniano_2012}.
\begin{equation}
    T_{\rm dust} \; = \; T_0 U^{(\frac{1}{4+\beta})}
\end{equation}

\noindent
where \( T_0 = 18.3\,\mathrm{K} \) represents the \(T_{\rm dust}\) in the solar neighborhood \citep{Mathis_1983}, and \(\beta = 1.79\) is the dust emissivity index adopted from the THEMIS model, which we fix at this value throughout the analysis. Physically, variations in \(\beta\) can reflect differences in the intrinsic properties of dust grains, such as their composition and structure, while statistically, variations in \(\beta\) influence the best-fit \(T_{\rm dust}\) values, with higher \(\beta\) generally leading to lower \(T_{\rm dust}\) estimates \citep{Kelly_2012}. This reflects an anti-correlation between the two parameters that can arise during the fitting process.

In this work, a single \( T_{\rm dust} \) value is assigned to each position within the galaxy, although in reality, \( T_{\rm dust} \) may vary within a resolution element due to multiple local heating sources.
Figure~\ref{figdust} shows the $\Sigma_{\rm dust}$ (left panel) and $T_{\rm dust}$ (right panel) maps derived for the galaxy NGC~3031 (M81). This figure shows that NGC~3031 exhibits a strong depletion of $\Sigma_{\rm dust}$ toward the galaxy center, while $T_{\rm dust}$ increases.
As NGC~3031 is situated in a dense environment and interacts with at least two neighboring galaxies \citep[M82 and NGC~3077, e.g.,][]{Kaufman_1989,Davidge_2008}, it is plausible that this galaxy was affected by recent encounters or merger events \citep[e.g.][]{Mattsson_Anderson_2012}. These events can substantially erode the dust, causing a central decline.
The low dust mass content in the central region of NGC~3031 is consistent with the very weak CO emission observed in the same region \citep[e.g.,][]{Solomon_1987,Brouillet_1988,Sakamoto_2001,Casasola_2007}. 
For this reason, NGC~3031 is a kind of prototype of CO-poor galaxies, 
as is Andromeda \citep[M31, e.g.,][]{Nieten_2006} or the center of NGC 7331 \citep[e.g.,][]{Tosaki_1997}, to a lesser extent.
The increase in $T_{\rm dust}$ toward the center of NGC~3031 could be due to different reasons, including
an AGN that is present in this galaxy (see Table~\ref{tab:sample}).

\begin{figure*}[h!]
    \centering
    \begin{minipage}{0.48\textwidth}
        \centering
        \includegraphics[width=\linewidth]{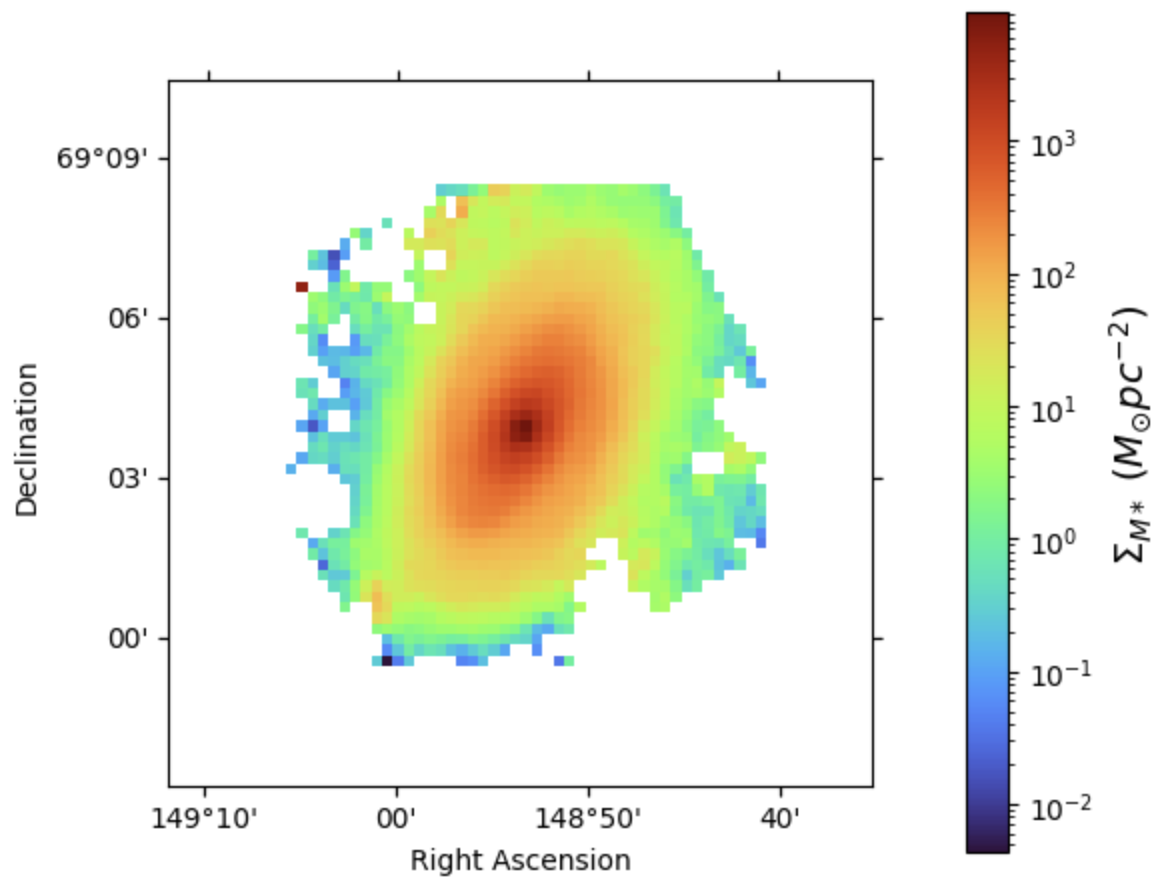}
    \end{minipage} 
    \hfill
    \begin{minipage}{0.48\textwidth}
        \centering
        \includegraphics[width=\linewidth]{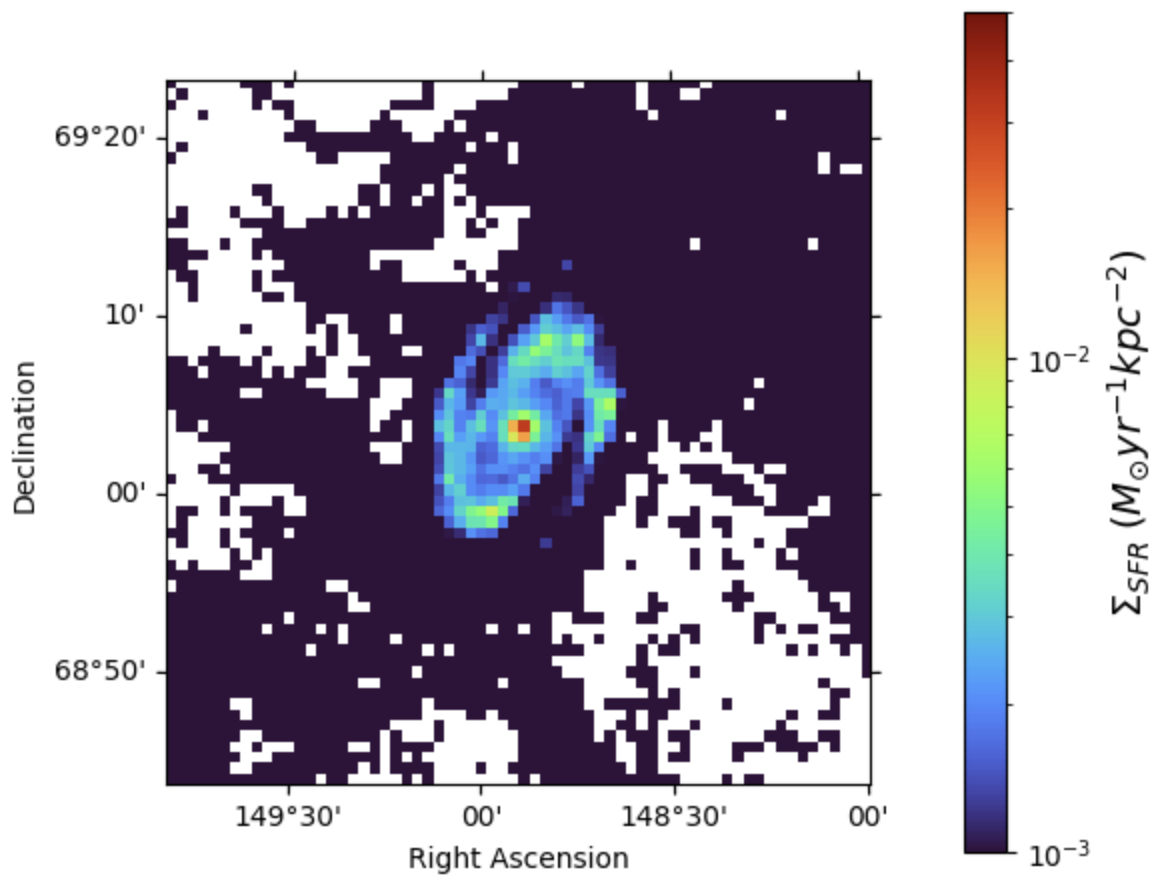}
    \end{minipage} 
    \caption{
    Same as Fig.~\ref{figdust} for $\Sigma_{M_*}$ (left) and $\Sigma_{\rm SFR}$ maps (right) of the sample galaxy NGC~3031 (M81) derived according to the prescriptions described in Sect.~\ref{sec:stellar-SFR}. Only regions with signal-to-noise ratio (S/N) > 2 are shown.
    }
    \label{figstellarmasssfr}
\end{figure*}

 \subsection{Stellar mass and star-formation rate surface density}
\label{sec:stellar-SFR}

Here, we recall the main steps used in the derivation of $\Sigma_{M_*}$ and $\Sigma_{\rm SFR}$ maps of C17.
The $\Sigma_{M_*}$ maps were derived using IRAC \(3.6 \, \mu\)m and \(4.5 \, \mu\)m images,
following the method of \cite{Querejeta_2015}, based on Independent Component Analysis \citep[ICA;][]{Meidt_2011}. 
In nearby galaxies, the old stellar population is the main source of the IRAC \(3.6 \, \mu\)m and \(4.5 \, \mu\)m fluxes, although additional contributions may occur.
For example, due to the finite bandwidth of the instrument, non-stellar features, such as the \(3.3 \, \mu\)m PAH emission feature and the PAH continuum, may contaminate these bands \citep{Flagey_2006}.
ICA enables us to separate the dominant stellar emission from these other contributions without prior assumptions about their relative strengths.
ICA provides a method for distinguishing between the dominant old stellar population and additional emissions without prior knowledge of the relative proportions of the sources. 
This technique results in a clean, smooth map of the old stellar light, consistent with expectations for an old, 
dust-free stellar population. 
For more details on implementing ICA to obtain stellar mass, refer to \cite{Querejeta_2015}.
 
The \(\Sigma_{\text{SFR}}\) maps were derived by combining the GALEX-FUV and WISE \(22 \, \mu\)m data, 
according to the calibration of \cite{Bigiel_2008}. Originally, \cite{Bigiel_2008}  used \textit{Spitzer} \(24 \, \mu\)m emission, instead of WISE \(22 \, \mu\)m. The GALEX-FUV data primarily trace O and early B stars, making it a good indicator of SF activity. However, FUV wavelengths are strongly affected by dust, which absorbs FUV light and re-radiates it at longer wavelengths in the mid-IR. To account for dust attenuation, the \(22 \, \mu\)m (or \(24 \, \mu\)m) flux densities are used to correct the FUV luminosity. The \(\Sigma_{\text{SFR}}\) was estimated from the GALEX–FUV emission corrected by the WISE \(22 \, \mu\)m following this calibration:

\begin{equation}
    \Sigma_{\text{SFR}} = 3.2 \times 10^{-3} \times I_{22} + 8.1 \times 10^{-2} \times I_{\text{FUV}},
\end{equation}
\noindent
where \(\Sigma_{\text{SFR}}\) is in units of \(M_{\odot} \, \text{yr}^{-1} \, \text{kpc}^{-2}\), 
and \(I_{22}\) and \(I_{\text{FUV}}\) are the \(22 \, \mu\)m and FUV intensities, respectively, in units of \(\text{MJy} \, \text{sr}^{-1}\). This calibration is based on the initial mass function (IMF) from \cite{Calzetti_2007}. 
This combined SFR calibration, together with others  \citep[e.g., H$\alpha$~+~\(24 \, \mu\)m,][]{Kennicutt_2007,Calzetti_2007}, is widely used for SFR studies of nearby galaxies 
\citep[e.g.,][]{Rahman_2011,Ford_2013,Momose_2013,Casasola_2017,Muraoka_2019,Yajima_2021,Casasola_2022}.
Figure~\ref{figstellarmasssfr} shows, as an example, the \(\Sigma_{M_*}\) (left panel) and \(\Sigma_{\rm SFR}\) (right panel) maps derived for the galaxy NGC~3031 (M81). 
For more information on the derivation of the $T_{\rm dust}$, $\Sigma_{\rm dust}$, \(\Sigma_{M_*}\), and \(\Sigma_{\rm SFR}\) maps, refer to C17.

 All the maps used in this analysis (\(\Sigma_{\rm dust}, T_{\rm dust}, \Sigma_{M_*}, \Sigma_{\rm SFR}\)) were originally convolved to a common angular resolution of 36\arcsec (the FWHM of \textit{SPIRE-500} \(\mu\)m beam) using PSF-matching convolution kernels from \cite{Aniano_2011}. 
The resulting maps were initially sampled at a pixel scale of 12\arcsec, corresponding to 1/3 of the SPIRE-500 beam size. 
However, to avoid artificial correlations arising from oversampling and to ensure that each pixel represents an approximately independent resolution element, we rebinned all maps to a final pixel size of 36\arcsec, matching the beam FWHM. 
This rebinned set of maps was used for all subsequent pixel-by-pixel correlation analyses.
In addition, all DustPedia images used here have been processed to remove Galactic foreground stars, following the approach described in \cite{Clark_2018} and C17. This was accomplished using the Python Toolkit for SKIRT \citep{Camps_2015, Verstocken_2017}, which performs automated stellar source removal. 

We note that one of the galaxies in our sample, NGC~5194 (also known as M~51a), belongs to an interacting system with its companion galaxy NGC~5195.
Consequently, the physical properties derived for NGC~5194 may be influenced by contamination from the stellar populations of 
NGC~5195. 
Surface brightness profiles from FUV to 500~$\mu$m presented in C17 show that this contamination mainly affects the older stellar component, while its impact on tracers of young stars is expected to be negligible. 
This view is reinforced by the 3D radiative transfer modeling of the NGC~5194–NGC~5195 pair by \citet{Nersesian_2020b}, 
who mapped the radiation fields of both galaxies and quantified the energy exchange within the system. 
Their model reveals that NGC~5195 contributes only 5.8\% to the total dust heating of the system. 
Nevertheless, in the regions of NGC~5194 close to NGC~5195, this influence is stronger, with the fraction of absorbed energy reaching 38\%. 
In contrast, on the outer disc of NGC~5194, the contribution from NGC~5195 falls below 1\%.

\begin{figure*}[htbp]
    \centering
    \includegraphics[width=\textwidth]{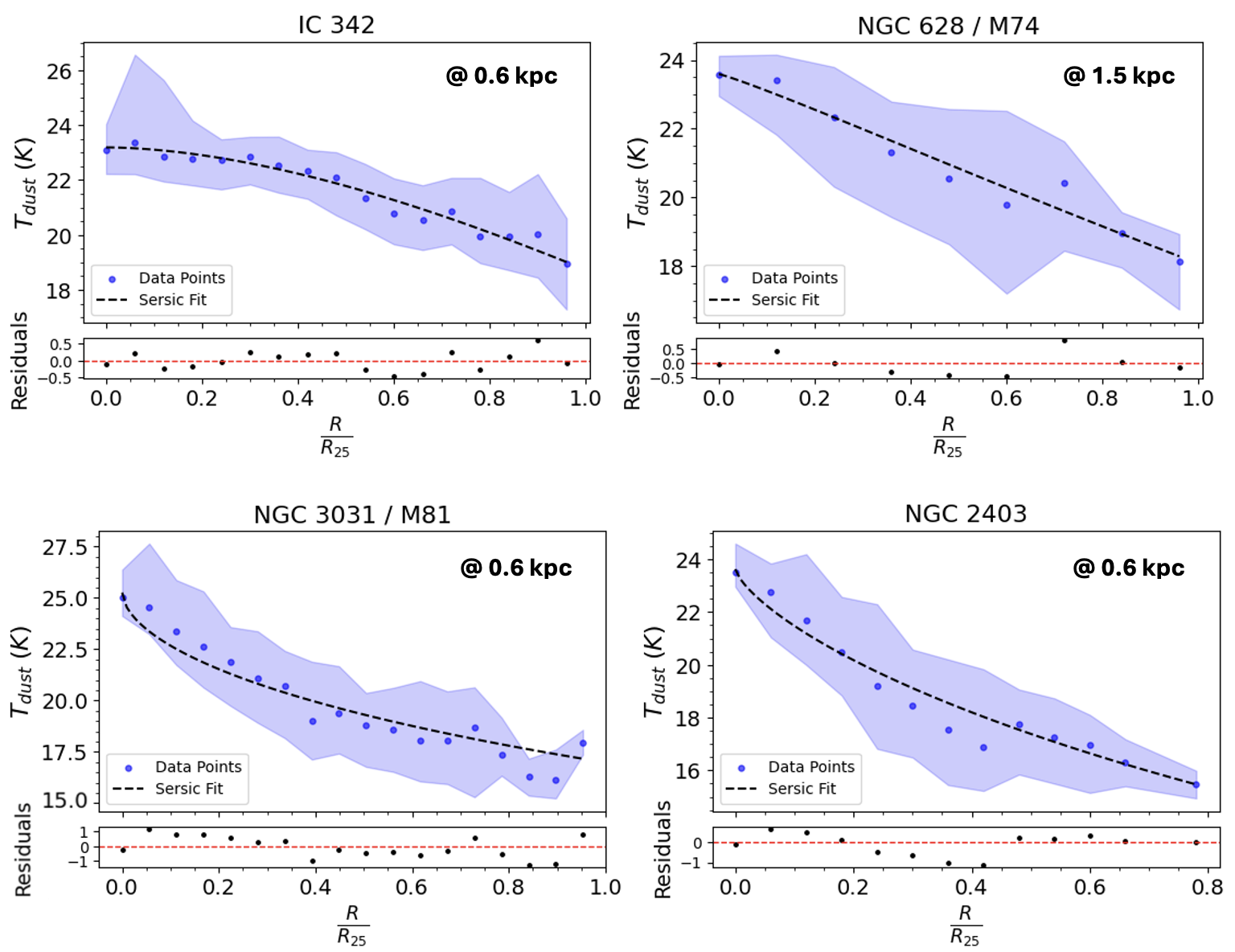}
    \caption{
         $T_{\rm dust}$ radial profiles as a function of the galaxy radius normalized by $R_{25}$ for some sample galaxies (IC~342, NGC~628, NGC~3031, NGC~2403). 
         Median $T_{\rm dust}$ values are shown as blue points, with the shaded blue regions indicating the total uncertainties, computed as the quadrature sum of the 16th and 84th percentiles and the modeling uncertainties (see Sect.~\ref{sec:radial profiles}).
         The dashed black line represents the Sérsic profile fit (see Eq.~\ref{eq:sersic} and Table~\ref{tabsersic}). 
         The sampled physical scale for each galaxy is quoted in its respective panel. 
         The lower panels display the residuals relative 
         to the Sérsic fit. 
    }
    \label{figradprof}
\end{figure*}

\section{Results}
\label{sec:results}
\subsection{Dust temperature radial profile}
\label{sec:radial profiles}
In this section, we examine the $T_{\rm dust}$ radial profiles for our galaxy sample. 
We construct these profiles using the $T_{\rm dust}$ maps presented in Sect.~\ref{sec:mass-t}. Each galaxy's map $T_{\rm dust}$ map is divided into concentric rings to extract the radial profile. We adopt a fixed radial binwidth equal to 36\arcsec, corresponding to the pixel scale of our maps. The number of radial bins varies depending on galaxy size, ranging from 9 bins for the smallest galaxies to 19 bins for the largest galaxies in our sample.  For each ring, the radial distance of each pixel is computed, and the $T_{\rm dust}$ values within the ring (between \(r_i\) and \(r_{i+1}\)) are used to determine the median $T_{\rm dust}$. Uncertainties in the median $T_{\rm dust}$ values are calculated as the quadrature sum of the 16th and 84th percentiles and the SED fitting uncertainties.
The average SED fitting uncertainties across all galaxies is $0.5$~K   , with a range spanning from $0.1 - 2.3~K$. The radial distance is normalized by the optical radius \(R_{25}\), enabling consistent comparison of profiles across galaxies of different sizes.

The $T_{\rm dust}$ radial profiles are fitted using a Sérsic function:
\begin{equation}
    T_{\rm dust}(r) = a \times \exp{(b  r^{1/c})} 
    \label{eq:sersic}
\end{equation}
\noindent where \(a\) is a scaling factor that represents the central {$T_{\rm dust}$, \(b\) quantifies the steepness of the $T_{\rm dust}$ gradient, and \(c\) determines the profile's curvature. The fitting is performed using the function \textit{curve\_fit} of \textit{Scipy} PYTHON module \citep{Scipy}.

\begin{table*}[ht!]
    \centering
    \caption{ Best-fit parameters of the Sérsic function applied to the $T_{\rm dust}$ radial profiles of the sample galaxies (see Eq.~\ref{eq:sersic}). Column 2 gives the physical scale sampled by our dataset.}
    \label{tabsersic}
    \begin{adjustbox}{max width=\textwidth}
    \begin{tabular}{lcccccl}
        \hline
        Galaxy & Physical scale & \(a\)    & \(b\) & \(c\) \\
        & [kpc]&&&\\
        \hline
        NGC 300         & 0.3 & \(20.94 \pm 0.29\)&\(-0.36 \pm 0.02\) &\(0.96 \pm 0.18\)\\
        NGC 2403          & 0.6 & \(23.62 \pm 0.27\) & \(-0.51 \pm 0.02\) & \(1.38 \pm 0.13\) \\
        IC 342        &  0.6 & \(23.18 \pm 0.16\) & \(-0.21 \pm 0.01\) & \(0.56 \pm 0.07\) \\
        NGC 7793        & 0.6 & \(21.47 \pm 0.18\) & \(-0.31 \pm 0.02\) & \(0.48 \pm 0.07\) \\
        NGC 3031 (M 81) & 0.6 & \(25.25 \pm 0.79\) & \(-0.40 \pm 0.04\) & \(1.77 \pm 0.41\) \\
        NGC 6946 & 0.8 & \(25.23 \pm 0.25\) & \(-0.35 \pm 0.01\) & \(0.62 \pm 0.06\) \\
        NGC 4736 (M 94)        & 0.8 & \(26.11 \pm 0.77\) & \(-0.48 \pm 0.06\) &  \(0.46 \pm 0.11\) \\
        NGC 5236 (M 83) & 0.8 & \(26.17 \pm 0.33\) & \(-0.37 \pm 0.02\) & \(0.61 \pm 0.10\) \\
        NGC 3621        & 1.2 & \(23.06 \pm 0.63\) & \(-0.27 \pm 0.03\) & \(1.11 \pm 0.23\) \\
        NGC 5457 (M 101)& 1.2 & \(22.97 \pm 0.97\) & \(-0.39 \pm 0.05\) & \(0.87 \pm 0.44\)\\
        NGC 5194 (M 51) & 1.3 & \(24.13 \pm 0.34\) & \(-0.30 \pm 0.04\) & \(0.35 \pm 0.10\) \\
        NGC 925         & 1.5 & \(20.43 \pm 0.17\) & \(-0.15 \pm 0.01\) & \(1.34 \pm 0.28\) \\
        NGC 628 (M 74)  & 1.5 & \(23.59 \pm 0.15\)& \(-0.27 \pm 0.02\) & \(0.90 \pm 0.13\) \\
        NGC 5055 (M 63) & 1.6 & \(22.85\pm 0.27\) & \(-0.28 \pm 0.02\) & \(1.34 \pm 0.24\) \\
        NGC 4725        & 2.2 & \(20.50 \pm 0.45\) & \(-0.29 \pm 0.03\) & \(0.61 \pm 0.16\) \\
        NGC 3521        & 2.3 & \(24.28 \pm 0.29\) & \(-0.34 \pm 0.01\) & \(0.95 \pm 0.08\) \\
        NGC 1097        & 2.7 & \(23.32 \pm 0.19\) & \(-0.41 \pm 0.02\) & \(1.13 \pm 0.16\) \\
        NGC 1365        & 3.0 & \(21.44 \pm 0.79\) & \(-1.39 \pm 0.61\) & \(0.14 \pm 0.05\) \\
        \hline
    \end{tabular}
    \end{adjustbox}
\end{table*}

The Sérsic fit parameters for all galaxies in our sample are summarized in Table \ref{tabsersic}. Examples of the individual $T_{\rm dust}$ radial profiles for galaxies IC~342, NGC~628 (M~74), NGC~3031 (M~81), and NGC~2403, along with their respective fits, are shown in Fig.~\ref{figradprof}. Figure~\ref{appfigradprof} displays {$T_{\rm dust}$ radial} profiles of the entire sample. To obtain a more general trend of how \(T_{\rm dust}\) decreases with radius, we compute a median \(T_{\rm dust}\) profile by combining the normalized radial profiles of all galaxies, sampled at the same interval. At each sampled radial point, the median $T_{\rm dust}$ across all galaxies is calculated, along with the 16th and 84th percentiles, to provide a measure of the scatter. This approach results in a representative $T_{\rm dust}$ radial profile for the entire sample, capturing the overall trend and variations. The average $T_{\rm dust}$ radial profile is shown in the panel $a$} of Fig.~\ref{figavprof}, and the best-fit Sérsic parameters for this median profile are presented in Table \ref{tabavgprof}. 
The average $T_{\rm dust}$ profile is well reproduced by the Sérsic function, with a reduced chi-squared \(\chi^2_{\rm red}\) value of $0.984$ and an associated $p$-value of $\leq 0.001$ indicating a very  good fit. 
The $T_{\rm dust}$ has a peak of $\sim$24~K in the galaxy center, and it decreases down to $\sim$15~K at a radius of approximately $R_{25}$. 

We also explore whether the presence of a bar influences the $T_{\rm dust}$ radial profiles by separating the barred galaxies (SAB-SB) and the unbarred galaxies and reproducing the median profiles, shown in the panels $b$ and $c$ of Fig.~\ref{figavprof}. 
We find no significant differences in the $T_{\rm dust}$ profiles due to the bar. Although we would have liked to examine the differences between interacting and non-interacting galaxies, nearly all galaxies in our sample are classified by NED\footnote{https://ned.ipac.caltech.edu/} 
as members of groups with varying degrees of interaction, which limits our ability to conduct this comparison.

\subsection{AGN contribution to temperature radial profile}
\label{sec:agn}
AGNs are known to significantly influence ISM through both radiative and mechanical feedback mechanisms (AGN-driven winds, outflows and jets), thereby affecting the thermal and chemical processes of various ISM components \citep[e.g.,][]{Fabian_2012,Kirkpatrick_2015,Viaene_2020,McKinney_2021}. In particular, the optical/UV AGN emission can be absorbed by the surrounding circum-nuclear dust, which is then heated, producing strong emissions in the mid-infrared (MIR) spectrum \citep[e.g.,][]{Fritz_2006,Pozzi_2012,Alonso-Herrero_2021}. How the AGN affects the colder and more diffuse dust component and enhances or decreases the FIR emission is still debatable, with studies on high-$z$ galaxies showing different conclusions \citep[e.g.,][]{Mullaney_2012,Santini_2012,Lanzuisi_2017}. In the local Universe, studies have explored the contribution of AGNs to dust heating, with findings varying based on AGN luminosity and other galaxy-specific factors. For example, \cite{Mullaney_2011} highlighted how moderate-luminosity AGNs can contribute significantly to dust infrared emission. While, \cite{Dickens_2022} showed that AGNs can heat dust at FIR wavelengths, with correlations observed between AGN power indicators and thermal emission from cool dust.
 
RT simulations studies suggest that AGNs significantly contribute to the heating of diffused dust in the host galaxy, particularly in regions close to the central AGN \citep[]{Duras_2017}. This role is most evident in the central regions, where AGNs can strongly influence $T_{\rm dust}$ profiles. For example, \cite{Schneider_2015} used RT simulations to investigate the IR properties of high-$z$ quasars ($z \sim 6$), demonstrating that AGN radiation can be a significant source of dust heating in the host galaxy, powering at least $30 \%$ and up to $70\%$ of the observed FIR emission. 

Although both LINERs and Seyfert galaxies fall under the AGN category, their effects on dust heating differ.
LINERs are characterized by low radiative efficiency and accretion rates, which result in weaker ionizing radiation compared to Seyferts
\citep[e.g.,][]{Macchetto_2006, Ho_2008}. 
This results in LINERs, typically associated with weaker ionizing radiation and older stellar populations, having minimal impact on dust heating, 
as they lack the intense UV and X-ray output required to significantly affect $T_{\rm dust}$ beyond the immediate nuclear regions.
In contrast, Seyferts exhibit stronger radiation fields and higher luminosities, which can elevate $T_{\rm dust}$ to a larger spatial extent.
Given this, our analysis will focus exclusively on Seyfert galaxies to assess the AGN-driven heating of cold dust. 
Thus, in the following discussion, ``AGN'' will refer only to Seyfert galaxies, while galaxies identified as LINERs or H{\sc ii} regions were grouped as ``No AGN''. Among our sample,  two galaxies have ambiguous nuclear classification. 
NGC~5236 classified as H\textsc{ii}/LINER, is included in the ``No AGN'' group, while 
NGC~3621, classified as LINER/Seyfert is included in the AGN category to evaluate possible AGN-related effects on dust heating.

\begin{figure}[h!]
    \centering
    \includegraphics[width=\columnwidth]{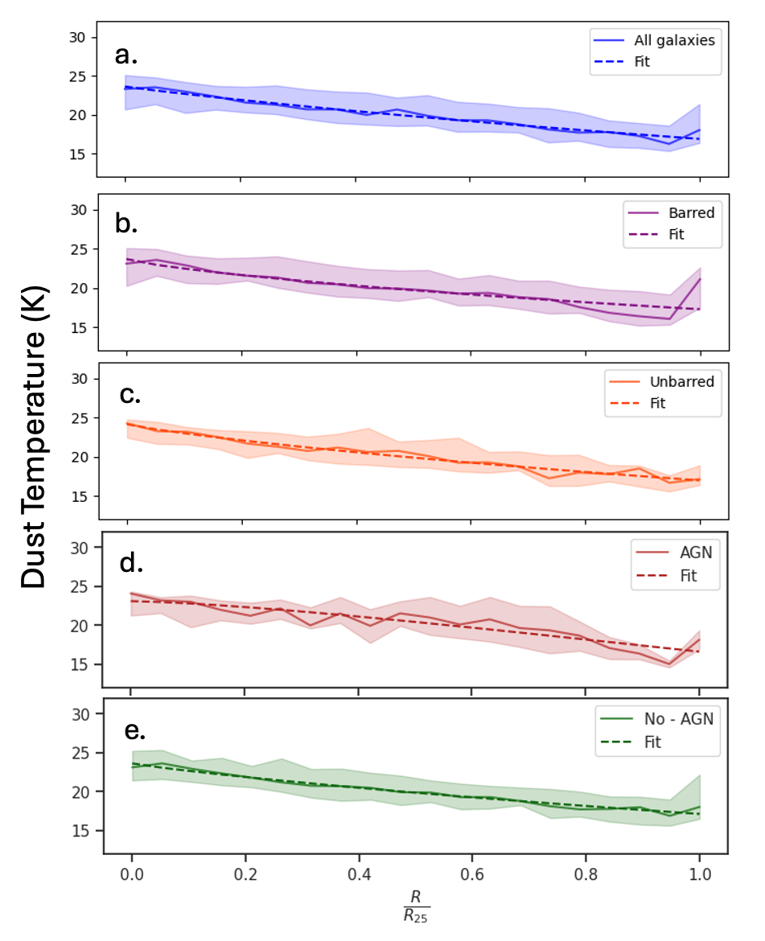}
    \caption{
 The panel $a$ shows the median $T_{\rm dust}$ radial profile for all galaxies in our sample (blue).  The panels $b$ and $c$ display the median $T_{\rm dust}$ profiles for barred (SAB-SB, purple) and unbarred (orange) galaxies, respectively. The panels $d$ and $e$ show the median $T_{\rm dust}$ profiles for AGN galaxies (red) and non-AGN (green) galaxies, respectively. The $T_{\rm dust}$ radial profiles are plotted as a function of the galaxy radius normalized by $R_{25}$. The dashed line represents the fitted Sérsic profile (see Eq.~\ref{eq:sersic}), with the corresponding parameters listed in Table~\ref{tabavgprof}. }
    
    \label{figavprof}
\end{figure}

To investigate whether AGNs contribute to dust heating in galaxies in our sample, we compared the average $T_{\rm dust}$ radial profiles of galaxies with and without AGNs. 
The results of this analysis are shown in the panels $d$ and $e$ of Fig.~\ref{figavprof}. 
We do not find significant difference between AGN and non-AGN galaxies.
This suggests that the contribution of AGNs to dust heating, if present, 
is not observable at the spatial resolution of our data, ranging from 0.3 to 3~kpc.
This also indicates that the influence of AGN on dust heating is restricted to very small spatial scales and therefore is likely to be diluted when averaged over the larger spatial regions investigated in this study. 
\begin{table*}[ht!]
    \centering
    \caption{
     Best-fit parameters of the S\'ersic function applied to the $T_{\rm dust}$ radial profiles for all sample galaxies, barred and unbarred galaxies, and AGN and non-AGN galaxies shown in Fig.~\ref{figavprof} (see Eq.~\ref{eq:sersic}).
     }
    \begin{adjustbox}{max width=\textwidth}
    \begin{tabular}{lccccl}
        \hline
         & \(a\)& \(b\) & \(c\) \\
        \hline
        All galaxies                & \(23.61 \pm 0.37\) & \(-0.34 \pm 0.02\) & \(1.07 \pm 0.14\) \\
        Barred galaxies (SAB-SB)    & \(23.68 \pm 0.97\) & \(-0.32 \pm 0.04\) & \(1.28 \pm 0.44\)\\
        Unbarred galaxies           & \(24.13 \pm 0.41\) & \(-0.35 \pm 0.02\) & \(1.16 \pm 0.15\)\\
        AGN galaxies                & \(23.08 \pm 0.64\) & \(-0.33 \pm 0.04\) & \(0.70 \pm 0.20\) \\
        Non-AGN galaxies            & \(23.60 \pm 0.31\) & \(-0.32 \pm 0.01\) & \(1.13 \pm 0.13\)\\
        \hline
    \end{tabular}
    \label{tabavgprof}
     \end{adjustbox}
\end{table*}

This finding is consistent with the results of \cite{Esposito_2022}, who also found no evidence for the influence of AGN on the cold and low-density molecular gas on the kpc scales in a sample of 35 local active galaxies.
Additionally, \cite{Viaene_2020}, through 3D RT modeling of the low-luminosity AGN hosted in NGC~1068, found that the contribution of the AGN to the heating of dust is negligible on global scales, with significant heating confined to the inner few hundred parsecs.

\subsection{Stellar dust heating mechanisms}
\label{sec:heating mech}
The stellar populations contribute to dust heating through various mechanisms. Three primary factors regulate the efficiency of dust grain heating: 1) the spatial distribution of stars within a specific population relative to the dust distribution in the galaxy; 2) the spectrum of different stellar populations, which determines their capacity to heat surrounding dust grains; 3) the topology of the ISM. These factors collectively shape the ISRF of the galaxy, which is responsible for dust grains heating. Understanding the interaction between the energy emitted by different stellar populations and dust grains is essential for understanding galaxy evolution.

To identify the primary heating mechanism of cold dust, we use two methods.
In the first method, we study the correlations of \(T_{\rm dust}\) with \(\Sigma_{\rm SFR}\) and \(\Sigma_{M_*}\), in the second one, we explore the relationship of \(T_{\rm dust}\) with \(\Sigma_{\rm dust}\) outlined by \cite{Utomo_2019}.

\subsubsection{Correlating dust temperature with star-formation rate and stellar mass}
\label{sectdsfrmstar}
\begin{figure*}[h!]
    \centering
    \includegraphics[width=1\linewidth]{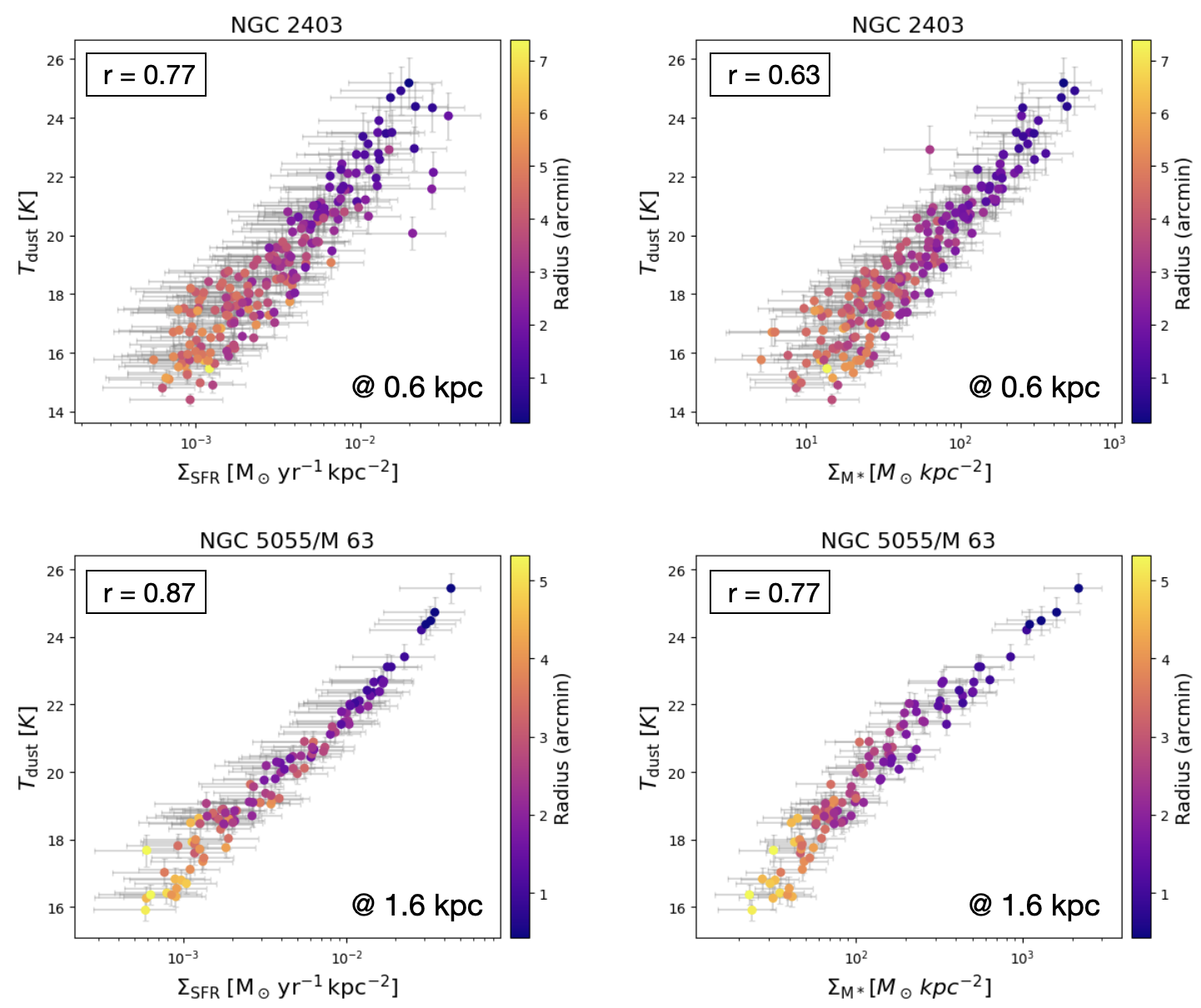}
    \caption{Examples of the pixel-by-pixel $T_{\rm dust}$--$\Sigma_{\rm SFR}$ (left panels) and $T_{\rm dust}$--$\Sigma_{M_*}$ (right panels) correlations for two sample galaxies (NGC~2403, NGC~5055). 
    The color bar represents the galaxy radius. 
    The weighted Pearson correlation coefficient $r$ (upper left corner) and the sampled physical scale (lower right corner) are quoted in the panels.}
    
        \label{figcorrexample}
\end{figure*}

Studying the correlations of $T_{\rm dust}$ with $\Sigma_{\rm SFR}$ and $\Sigma_{M_*}$ is a tool to understand the link between $T_{\rm dust}$ and different stellar populations.
Young, massive stars are the main contributors to the heating of dust in active star-forming regions and can be traced by $\Sigma_{\rm SFR}$. These stars have a characteristic timescale of about 100 Myr, which is the timescale traced by FUV emission. In contrast, evolved stars contribute to the heating of dust in more quiescent regions and can be traced by $\Sigma_{M_*}$.
Once we derived $T_{\rm dust}$, $\Sigma_{\rm SFR}$, and $\Sigma_{M_*}$ maps for all sample galaxies according to the prescriptions presented in Sect.~\ref{sec:sample}, we study the pixel-by-pixel $T_{\rm dust}$--$\Sigma_{\rm SFR}$ and $T_{\rm dust}$--$\Sigma_{M_*}$ correlations using weighted Pearson correlation coefficients.
We focus on the region within the optical disk defined by \(R_{25}\). 
This choice is a normalization parameter already used in the literature \citep[e.g., C17,][]{Casasola_2020,Enia_2020,Morselli_2020,Casasola_2022}. 

Figure \ref{figcorrexample} shows an example of the $T_{\rm dust}$--$\Sigma_{\rm SFR}$ and $T_{\rm dust}$--$\Sigma_{M_*}$ correlations for two sample galaxies (NGC~2403, NGC~5055), and Table~\ref{Corr values} collects the weighted Pearson correlation coefficients\footnote{Typically, if \(0 < |r| < 0.3\), the correlation is defined as weak; if \(0.3 \leq |r| < 0.7\), as moderate; and if \(|r| \geq 0.7\), as strong. Additionally, one common interpretation of the correlation coefficient \(r\) is that its square, \(r^2\), represents the fraction of the variation in one variable that can be explained by the other. For example, an \(r\) value of 0.50 between \(T_{\mathrm{dust}}\) and \(\Sigma_{\mathrm{SFR}}\) means that 25\% of the variation in \(T_{\mathrm{dust}}\) can be explained by \(\Sigma_{\mathrm{SFR}}\).
}, $r$, of the correlations for the entire sample. 
The $r$ coefficients range from moderate to high values for both correlations, indicating that both SF activity and old stars contribute to dust heating. 
However, our analysis shows that in nearly half of the sample (56$\%$, 10 out of 18 galaxies), the young stellar population dominates the dust heating process, as indicated by a higher Pearson correlation coefficient for the $T_{\rm dust}$--$\Sigma_{\rm SFR}$ relation compared to the $T_{\rm dust}$--$\Sigma_{M_*}$ relation. 
In the remaining galaxies, the evolved stellar population is the dominant contributor.

\begin{table}[h!]
    \centering
    \caption{Correlation coefficients and $p$-values for each sample galaxy.}
    \begin{adjustbox}{max width=\textwidth}
    \begin{tabular}{lcccl}
        \hline
        Galaxy & \(r_{T_{\rm dust}-\Sigma_{\rm SFR}}\) & \(r_{T_{\rm dust}-\Sigma_{M_*}}\) & \(p_{T_{\rm dust}-\Sigma_{\rm dust}}\)  \\
        \hline
        NGC 300             & 0.79 & 0.61 & 0.008\\
        NGC 2403            & 0.77 & 0.63 &0.016\\
        IC 342              & 0.36 & 0.71 & 0.963 \\
        NGC 7793            & 0.88 &  0.63 &0.001\\
        NGC 3031 (M 81)     & 0.64 & 0.66 &0.297\\
        NGC 6946            & 0.57 & 0.79 &0.994\\
        NGC 4736 (M 94)     & 0.70 & 0.73 &0.995\\
        NGC 5236 (M 83)     & 0.50 & 0.52 &0.981\\
        NGC 3621            & 0.81 & 0.60 &0.014\\
        NGC 5457 (M 101)    & 0.68 &  0.18 &0.159\\
        NGC 5194 (M 51)     & 0.67 & 0.70 &1.000\\
        NGC 925             & 0.83 & 0.37 &0.012\\
        NGC 628 (M 74)      & 0.85 & 0.70 & 0.254\\
        NGC 5055 (M 63)     & 0.87 &  0.77 &1.000\\
        NGC 4725            & 0.80 & 0.76  &0.239\\
        NGC 3521            & 0.79 & 0.69 &0.996\\
        NGC 1097            & 0.59 & 0.71 &0.193\\
        NGC 1365            & 0.40 &  0.62 &1.000\\
        \hline
    \end{tabular}
    \end{adjustbox}
    \tablefoot{
    Weighted Pearson correlation coefficients of the $T_{\rm dust}-\Sigma_{\rm SFR}$ \((r_{T_{dust}-\Sigma_{SFR}}\)) and ${T_{\rm dust}-\Sigma_{M_*}}$ ($r_{T_{\rm dust}-\Sigma_{M_*}}$) correlations (see Sect.~\ref{sectdsfrmstar}) and $p$-value of the \(\chi^2\) fit for the ${T_{\rm dust}-\Sigma_{\rm dust}}$ correlation for each sample galaxy (see Sect.~\ref{sectprobing}).
    \label{Corr values}}
\end{table}

\begin{figure*}[h!]
    \centering
    \includegraphics[width=1\linewidth]{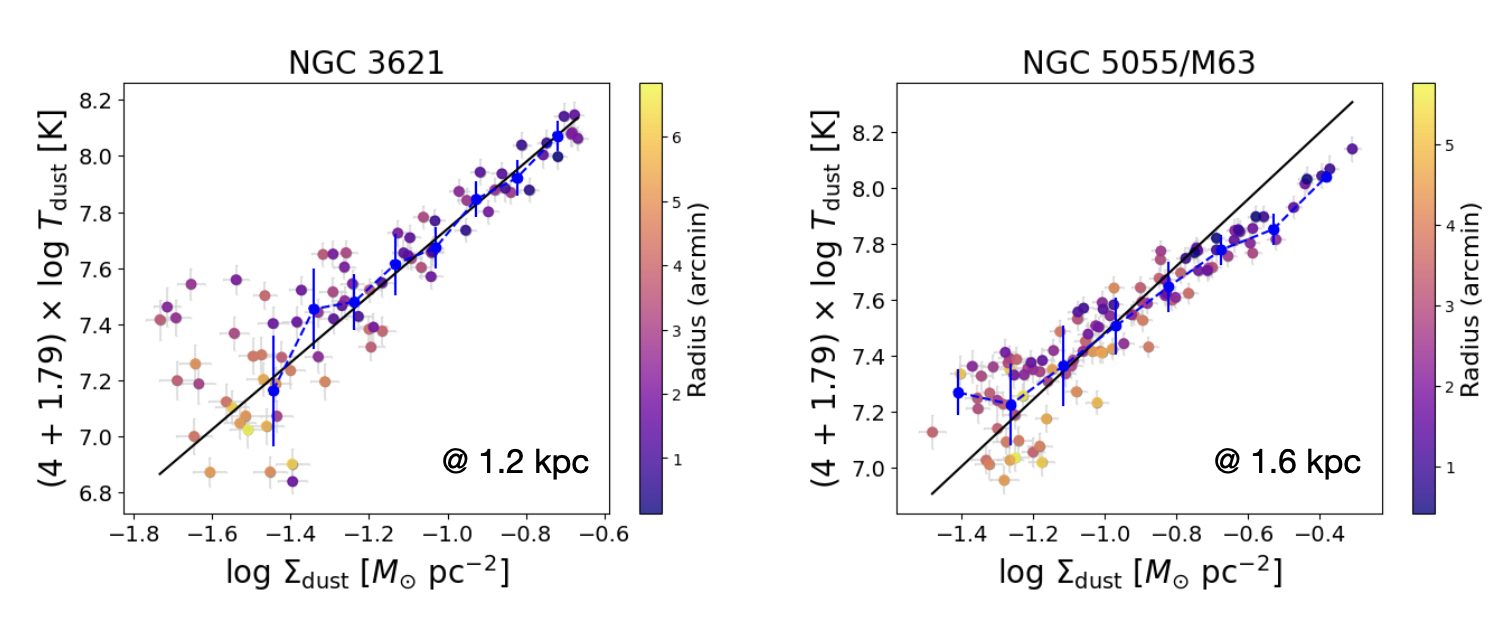}
    \caption{
 The pixel-by-pixel relation between \(\log T_{\rm dust}\) and \(\log \Sigma_{\rm dust}\) according to Eq.~(\ref{eq:temp_dust_final}) for two sample galaxy (NGC~3621, NGC~5055). 
    Blue points represent the mean values within each bin, with error bars indicating the standard deviation in each bin. The black line corresponds to the expected relation (see Eq. \ref{eq:temp_dust_final}). The color bar represents the galaxy radius.  The sampled physical scale for each galaxy is quoted in its respective panel.}
    \label{T-dustmass}
\end{figure*}

\subsubsection{Probing dust heating with \(\Sigma_{\rm dust}\) and \(T_{\rm dust}\)}
\label{sectprobing}
In this section, we investigate the contribution of young stellar populations to dust heating following the method developed by \cite{Utomo_2019}. 
In this method, a precise correlation between \(\Sigma_{\rm dust}\) and \(T_{\rm dust}\) is expected under the hypothesis that young stars are the dominant dust heating mechanism.
Here we summarize the main steps of the method. The main assumption of this method is that the entire IR luminosity ($L_{\rm IR}$) arises from the reprocessed radiation of young stars and therefore $L_{\rm IR}$ is a tracer of SFR. This implies that $L_{\rm IR}$ surface density ($\Sigma_{L_{\rm IR}}$) can be linked to $\Sigma_{\rm SFR}$ as $\Sigma_{L_{\rm IR}} \propto  \Sigma_{\rm SFR}$.
By using the Kennicutt-Schmidt (KS) relation \citep[][]{Schmidt_1959,Schmidt_1963,Kennicutt_1998a,Kennicutt_1998b}, 
i.e. the correlation between $\Sigma_{\rm SFR}$ and gas mass surface density 
($\Sigma_{\rm gas}$, $\Sigma_{\rm SFR} \propto \Sigma_{\rm gas}^n$), and assuming a linear relationship between $\Sigma_{\rm dust}$ and $\Sigma_{\rm gas}$ ($\Sigma_{\rm gas} \propto \Sigma_{\rm dust}$),
one can write:
\begin{equation}
\label{eq_lir_dust}
\Sigma_{L_{\rm IR}} \propto \Sigma_{\rm dust}^n
\end{equation}
\noindent 
In addition, \citet{Utomo_2019} assume that the dust is in thermal equilibrium with the radiation, described by a Modified Black Body (MBB) function, and it is optically thin. This implies that $L_{\rm IR}$ can be expressed as:
\begin{equation}
L_{\rm IR}= 4\pi M_{\rm dust} \int \kappa_{\nu} B_{\nu}(T_{\rm dust})\,d\nu \\
\end{equation} 
\noindent 
where $\kappa_{\nu}$ is the dust absorption cross section per unit mass. 
Following a standard approach, $\kappa_{\nu}$ is assumed to be a power law \citep{Draine_2011}, $\kappa_{\nu}= \kappa_0\left(\frac{\nu}{\nu_0}\right)^{\beta}$, where $\beta$ is the emissivity index and \(\kappa_0\) is the opacity at reference frequency $\nu_0$. 
By performing the integral in frequency, this yields: 
\begin{equation}
\label{eq_lir_t}
L_{\rm IR} \propto \kappa_0 M_{\rm dust} T_{\rm dust}^{4 + \beta}
\end{equation} 
\noindent
Equation~(\ref{eq_lir_t}) can be expressed in terms of logarithm as:
\begin{equation}
\label{eq:temp_dust_final}
(4+\beta) \log{T_{\rm dust}} = A + (n-1)\log {\Sigma_{\rm dust}},
\end{equation}
\noindent where \(A\) is a constant that encapsulates all proportionality constants, including geometrical factors.
\noindent We use Eq.~(\ref{eq:temp_dust_final}) by assuming \(\beta = 1.79\) (see Sect.~\ref{sec:mass-t}), $n = 2.19$ derived by \citet{Casasola_2022} studying the pixel-by-pixel KS relation for galaxies of our sample at the common resolution of 3.4~kpc, and the constant CO-to-H$_2$ conversion factor $X_{\rm CO} = N(H_2)/I_{\rm CO} = 2 \times 10^{20}$~cm$^{-2}$~(K~km~s$^{-1}$) derived by \citet{Bolatto_2013}\footnote{The CO-to-H$_{2}$ conversion factor expressed in terms of $\alpha_{CO}$ corresponds to $\alpha_{CO} = {\rm M(H_{2})}/{\rm L_{CO}} = 3.2\,{\rm M}_\odot\,{\rm pc}^{-2}$~(K~km~s$^{-1}$)$^{-1}$ \citep[][]{Narayanan_2012}.}.

We test the Utomo method by producing the pixel-by-pixel correlation
between log$T_{\rm dust}$ and log$\Sigma_{\rm dust}$ for each sample galaxy and by fitting it with Eq.~(\ref{eq:temp_dust_final}).
Any deviation of the observed correlations from the trend expected by 
the Utomo method may suggest that $L_{\rm IR}$ does not completely arise from the reprocessed radiation of young stars but other sources could contribute, such as older stellar populations and/or AGN activity.

To ensure the robustness of our results and to minimize the effects of correlated uncertainties, we restrict our analysis to pixels with a signal-to-noise ratio (S/N) greater than 5 for both the $T_{\rm dust}$ and $\Sigma_{\rm dust}$ maps. At lower S/N, uncertainties in the simultaneous derivation of \(T_{\rm dust}\) and \(\Sigma_{\rm dust}\) can become correlated, leading to an apparent anti-correlation between the two parameters. This occurs because a higher \(T_{\rm dust}\) can be compensated by a lower dust mass to match the observed emission. 
Moreover, we also impose a threshold on $\Sigma_{\rm dust}$, taking into account only pixels with log $\Sigma_{\rm dust}\rm{[M_{\odot}~pc^{-2}]} > -1.5$. This threshold was chosen to exclude regions where too low values of $\Sigma_{\rm dust}$ could introduce additional uncertainties.
 
For each sample galaxy, we perform a chi-square \((\chi^{2})\) fitting of the observed data versus the expected values from the relation between log$T_{\rm dust}$ and log$\Sigma_{\rm dust}$. 
The key indicator in this \(\chi^2 - \)test is the $p$-value, which quantifies how likely it is that our data align with Eq.~(\ref{eq:temp_dust_final}) by chance. 
We adopt a significance level of \(\alpha = 0.05\). 
For a given galaxy, a $p$-value below \(\alpha\) indicates that the observed relationship between $T_{\rm dust}$ and $\Sigma_{\rm dust}$ aligns with the trend expected by \citet{Utomo_2019}, and therefore the dust heating is mainly due to young stellar populations.
In contrast, a $p$-value exceeding \(\alpha\) implies that the assumption of dust heating mainly due to young stars does not hold, but other sources (e.g., older stars) contribute to it in a not negligible way.

Figure~\ref{T-dustmass} presents the results of our analysis for two representative galaxies in the sample 
(NGC~3621 and NGC~5055), while Fig.\ref{appT-dustmass} shows the corresponding results for the remaining galaxies.
These two cases illustrate the diversity within our sample: NGC~3621, with a $p$-value of 0.014, 
exhibits $T_{\rm dust}$ and $\Sigma_{\rm dust}$ values that closely follow the trend reported by \citet{Utomo_2019}, whereas NGC~5055, with a $p$-value of 1.000, significantly deviates from that relation.
In our sample of 18 galaxies, approximately 72\% (13 galaxies) have a $p$-value exceeding the significance threshold of \(\alpha = 0.05\), indicating that young stars are not the sole contributors to dust heating in these systems. 
Instead, other sources, most likely evolved stellar populations, appear to play a significant role in powering the observed IR emission. 
These results support a scenario in which the heating of cold dust is not exclusively linked to recent SF, but is also shaped by other processes that reflect the specific physical conditions of each galaxy.

\section{Discussion}
\label{sec:discussion}
\begin{figure}
    \centering
    \includegraphics[width=\columnwidth]{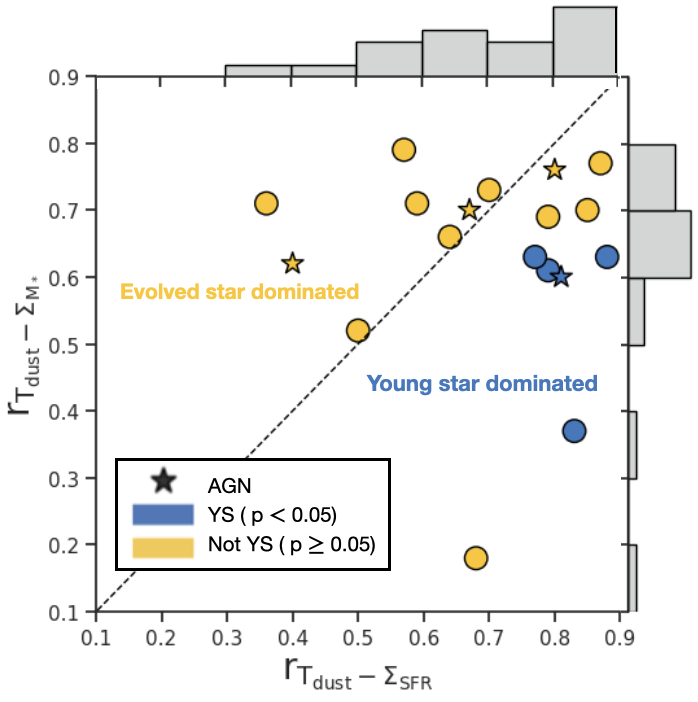}
    \caption{Pearson correlation coefficient $r$ of the \(T_{\rm dust} - \Sigma_{\rm SFR}\) relation  vs. that of the \(T_{\rm dust} - \Sigma_{\rm M_*}\) relation.
    AGN galaxies are marked with stars, non-AGN galaxies with circles.
    The galaxies are color-coded according to the $p$-values of the \(\chi^2\) test (see Sect.~\ref{sectprobing}): blue symbols correspond to $p < 0.05$ (aligned with the young star-driven heating assumption, YS), orange symbols to $p \geq 0.05$ (deviation from the young star heating assumption, not YS).
    The black dashed line represents the 1:1 relation. The grey histograms in the margins show the distributions of the correlation coefficients for \(T_{\rm dust} - \Sigma_{\rm SFR}\) (top) and \(T_{\rm dust} - \Sigma_{\rm M_*}\) (right). }
    \label{final}
\end{figure}
From our analysis, it emerges that both young and evolved stellar populations play significant roles in dust heating, though their contributions vary across galaxies. 
The plot in Fig.~\ref{final} illustrates this, with the $x$-axis representing the correlation coefficient between \(T_{\rm dust}\) and \(\Sigma_{\rm SFR}\) and the $y$-axis between \(T_{\rm dust}\) and \(\Sigma_{M_*}\).  
The black dashed reference line divides the plot into regions where \(T_{\rm dust}\) is more strongly influenced by either young stars (traced by $\Sigma_{\rm SFR}$) or evolved stars (traced by $\Sigma_{M_*}$). 
Galaxies above the line have a stronger correlation between \(T_{\rm dust}\) and $\Sigma_{M_*}$, where evolved stars likely dominate dust heating. 
In contrast, galaxies below the line exhibit a stronger correlation with SFR, suggesting that young stars are the dominant source of dust heating. 
However, we note that points located near the dashed line may correspond to cases where the first method provides a less reliable separation of the heating contributions.

In Fig.~\ref{final}, the symbols are color-coded based on the results of the second method ($T_{\rm dust}$--$\Sigma_{\rm dust}$ correlation).
Table~\ref{tab:heating_analysis} compares the dominant dust heating mechanism determined by the two methods based on $r$ and $p$-value for the first and second methods, respectively.

\begin{table}[h!]
    \caption{
    Dust heating mechanisms derived from both methods for each sample galaxy. 
    "YS" = young stars; "ES" = evolved stars. 
    AGN hosts are indicated by an asterisk (*).
    }
    \centering
    \begin{tabular}{lcc}
        \toprule
        Galaxy Name & Method 1 ($r$) & Method 2 ($p$-value)\\
        \midrule
        NGC~300             &   YS          &   YS  \\
        NGC~2403            &   YS         &   YS  \\
        IC~342              &   ES          &   ES  \\
        NGC~7793            &   YS          &   YS  \\
        NGC~3031 (M 81)    &   ES          &   ES  \\
        NGC~6946           &   ES          &   ES  \\
        NGC~4736 (M 94)   &   ES          &   ES  \\
        NGC~5236 (M 83)     &   ES          &   ES  \\
        NGC~3621*            &   YS          &   YS  \\
        NGC~5457 (M 101)   &   YS          &   ES  \\
        NGC~5194 (M 51)*    &   ES          &   ES  \\
        NGC~925             &   YS          &   YS  \\ 
        NGC~628 (M 74)      &   YS          &   ES  \\
        NGC~5055 (M 63)     &   YS          &   ES  \\
        NGC~4725*           &   YS          &   ES  \\
        NGC~3521            &   YS          &   ES  \\
        NGC~1097           &   ES          &   ES  \\
        NGC~1365*           &   ES          &   ES  \\
        \bottomrule
    \end{tabular}
    \label{tab:heating_analysis}
\end{table}

Out of the 18 galaxies analyzed, the two methods yield consistent results in 13 cases ($\sim$72\%). While the overall agreement is good, discrepancies are found in the following galaxies: 
NGC~5457, NGC~628, NGC~5055, NGC~4725, NGC~3521.
These disagreements may stem from various combined reasons. For instance, discrepancies could be due to resolution effects. Coarse resolution may indeed blend SFRs with the diffused ISM, leading to an averaging effect that obscures localized heating mechanisms. 
Additionally, we assume a constant slope (\(n = 2.19\), see Sect.~\ref{sectprobing}) for the KS law across all galaxies \citep{Casasola_2022}. 
Although this slope value has been derived for the same galaxy sample at the common resolution of 3.4~kpc, \citet{Casasola_2022} also showed that the KS slope varies galaxy-by-galaxy and as a function of resolution.
This suggests that specific physical conditions and SF efficiencies could influence the KS relation, and a single slope may not fully capture these variations.
Importantly, the adopted slope of \textit{n} = 2.19 itself relies on assumptions about the gas mass, including the use of a constant CO-to-H₂ conversion factor when estimating molecular gas from CO emission. Variations in this factor can directly affect the derived KS slope, introducing additional uncertainties. It is well known that the CO-to-H$_2$ conversion factor varies across galaxies and within galaxies \citep[e.g.,][]{Bolatto_2013}.
Studies like \cite{R_my_Ruyer_2013} observed a trend of color temperature with metallicity, indicating that low-metallicity (12~+~log(O/H)~$\lesssim$~8.3) systems\footnote{We point out that our sample is characterized by ``normal'' metallicity, with values 8.3~$\le$~12~+~log(O/H)~$\le$~8.8, based on the empirical metallicity calibration N2 of \citet{Pettini_2004} \citep[see][for details on metallicity of the sample.]{Casasola_2022}} tend to have hotter $T_{\rm dust}$, likely due to a stronger contribution from young star heating. 
In these low-metallicity environments, the CO-to-H$_2$ conversion factor may be higher, potentially leading to a different distribution of dust heating mechanisms compared to higher metallicity galaxies. 
The assumption of a constant the CO-to-H$_2$ conversion factor could therefore mask the metallicity-dependent variations in heating, contributing to some discrepancies observed in our analysis.
Another possible source of discrepancy between the two methods is the assumption of a linear relationship between \(\Sigma_{\rm dust}\) and \(\Sigma_{\rm gas}\) (see Sect.~\ref{sectprobing}). 
Variations in the dust-to-gas mass ratio (DGR) are observed and are mainly due to changes in gas-phase metallicity, grain growth, or destruction mechanisms.

Our findings align with previous studies, such as \cite{Nersesian_2019}, which used the CIGALE energy balance code to investigate dust heating sources on a global scale. Their analysis of \textit{Herschel} data from the DustPedia project revealed that in spiral galaxies, young stars contribute up to $\sim$77$\%$ of their luminosity to dust heating, while older stars contribute to $\sim$24$\%$.  
While these results reflect the dominant role of young stars in global heating, the contribution from evolved stellar populations is still significant.
Recent findings by \cite{Pas_2023} reinforce this trend, providing a detailed breakdown of dust heating contributions in different galaxy types using the GAMA survey. Their study found that star-forming Sa-Scd galaxies exhibit a dominant contribution from young stars, with $59\%$ of their luminosity used for dust heating, while older stars contribute only $16\%$. They found similar contributions even in the case of quiescent Sa-Scd, suggesting that residual SF continues to play a role.
\cite{Viaene_2017} studied the Andromeda galaxy (M31) to examine dust heating mechanisms, finding that the majority of dust heating ($\sim$91$\%$) of the absorbed luminosity is driven by evolved stellar populations. 
This finding further supports the idea that older stellar populations play an important role in dust heating on a galactic scale.

Several galaxies from our sample have RT models performed in other studies:
M51 has been analyzed by \cite{DeLooze_2014}, \cite{Nersesian_2020b}, and \cite{Inman_2023}; M81 by \cite{Verstocken_2020}, NGC~1365 by \cite{Nersesian_2020a}, NGC~628 by \cite{Rushton_2022}, and M101 by \cite{Pricopi_2025}. 
These studies found that young stars contribute to dust-heating between $\sim$50$\%$ and $\sim$75$\%$ of the total heating. 
For instance, in M51, different RT models estimate the fraction of dust heating from young stars to range from $63 \%$ to $72 \%$, while in M81, the heating is nearly evenly split between young and old stellar populations. 
These findings highlight that galaxy morphology, stellar content, and variations in SF activity can significantly influence the relative contributions of young and evolved stars to dust heating, underlining the importance of considering these factors when interpreting dust heating on galactic scales.

Our findings also support recent results from \cite{chastenet_2024}. 
Although their integrated measurements showed no correlation between the average radiation field intensity and stellar mass, their resolved analysis revealed a strong correlation between $U_{\rm min}$ and $\Sigma_{M_*}$ in a sample of $\sim$800 nearby galaxies. 
Their findings suggest that evolved stars maintain a consistent radiation field across the galaxy disk, contributing significantly to dust heating. This is further corroborated by studies from \cite{Bendo_2012,Bendo_2015} and \cite{Bonato_2024}, 
who similarly observed that young stars are primarily responsible for heating in active star-forming regions, whereas older stars offer more consistent background heating. 
The recent work by \cite{Chiang_2023} further investigates the relationship between
$T_{\rm dust}$ and various galaxy properties for a sample of 46 nearby galaxies. 
They found a strong correlation between $T_{\rm dust}$ and \(\Sigma_{\rm SFR}\), emphasizing the role of young stars in dust heating.
Furthermore, they found that for galaxies with similar \(\Sigma_{\rm SFR}\), an increased DGR leads to a lower $T_{\rm dust}$, likely due to enhanced shielding of dust in regions with a higher dust mass. 
Collectively, these findings emphasize the dual role of young and evolved stars: while young stars dominate localized heating in star-forming regions, older stars provide a baseline heating effect that is consistent and widespread.

 \section{Conclusions}
 \label{sec:conclusions}
In this study, we analyzed a sample of 18 large, nearby, face-on spiral galaxies from the DustPedia project to identify the dominant mechanisms for heating the cold dust component, traced by $\textit{Herschel}$, in these systems. 
We compare two methods. In the first method, we study the pixel-by-pixel correlations between $T_{\rm dust}$ and $\Sigma_{\rm SFR}$ (tracer of young stars) and between $T_{\rm dust}$ and $\Sigma_{M_*}$ (tracer of old stars). 
In the second method, we analyze the pixel-by-pixel correlation between $T_{\rm dust}$ and $\Sigma_{\rm dust}$. 
Our data sample physical scales from 0.3 to 3~kpc.

Our main results are as follows:
\begin{itemize}

    \item The $T_{\rm dust}$ radial profiles show a decline from $\sim$24~K at the galaxy center to $\sim$15~K at $R_{25}$, which is well described by the Sérsic function.
    \item The presence of a central AGN does not significantly influence $T_{\rm dust}$ radial profiles on the spatial scales sampled by our data.
    \item The analysis of the correlations between $T_{\rm dust}$ and $\Sigma_{\rm SFR}$ and between $T_{\rm dust}$ and $\Sigma_{M_*}$ (first method) suggests that both young stars and evolved stars contribute to dust heating. 
    In 56$\%$ of the galaxies, $T_{\rm dust}$ shows a stronger correlation with $\Sigma_{\rm SFR}$, indicating the dominant role of young stars in dust heating in star-forming regions. 
    \item The analysis of the correlation between \(T_{\rm dust}\) and \(\Sigma_{\rm dust}\) (second method) shows that for the $\sim$72\%  of the sample young stars are not the unique contributors to dust heating, and other sources (e.g., evolved stars) can play a role.
    \item Out of the 18 galaxies analyzed, the two methods are consistent in 13 cases ($\sim$72$\%$).
    The discrepancies between the two methods may be due to different combined factors, including resolution effects, the assumption of a constant value for the slope of the KS relation and for the CO-to-H$_2$ conversion factor, and the assumption of a linear relationship between $\Sigma_{\rm dust}$ and $\Sigma_{\rm gas}$.

\end{itemize}

Our study strongly outlines the need for statistical samples of galaxies with the SED properly sampled in the FIR–sub-mm part of the spectrum, for the purpose of an advanced characterization of the dust temperature and dust mass. This will be achieved by exploiting the synergies between sub-mm/mm (i.e. ALMA, NOEMA) and future space FIR facilities such as the forthcoming PRIMA telescope (PI: G.J. Glenn)\footnote{http://prima.ipac.caltech.edu}. 
Moreover, the gas-phase metallicity is absent in our analysis, although it represents an important element of the ISM.
However, the Metal-THINGS survey \citep{Lara-Lopez_2021} is producing resolved metallicity information for most galaxies of our sample which will allow us to take into account the variation of the metallicity within galaxy disks and its effect on dust and gas. This will be the subject of an incoming paper (Tailor et al. in prep.)

\begin{acknowledgements}
The authors thank the anonymous referee for his/her constructive comments and valuable suggestions, which helped improve the quality and clarity of the manuscript. We thank Fabrizio Gentile for his input and suggestions regarding the data analysis and statistical methods.
This research has made use of the NASA/IPAC Extragalactic Database (NED), 
which is funded by the National Aeronautics and Space Administration and operated by the California Institute of Technology. 
VC, FP, FC, and SB acknowledge funding from the INAF Mini Grant 2022 program ``Face-to-Face with the Local Universe: ISM’s Empowerment (LOCAL)''.
F.Galliano acknowledges support by the French National Research Agency under the contract WIDEN- ING ANR-23-ESDIR-0004, as well as by the Programme National “Physique et Chimie du Milieu In- terstellaire” (PCMI) of CNRS/INSU, with INC and INP, and the “Programme National Cosmologie et Galaxies” (PNCG) of CNRS/INSU, with INP and IN2P3, both programs being co-funded by CEA and CNES.
MR acknowledges support from project PID2023-150178NB-I00 (and PID2020-114414GB-I00), financed by MCIU/AEI/10.13039/501100011033, and by FEDER, UE.

\end{acknowledgements}

\bibliography{ref}
\clearpage
\onecolumn
\appendix

\section{Images of Radial Profiles}
\label{appendix: radial prof}
\begin{figure}[H]
\centering
\begin{tabular}{ccc}
\includegraphics[width=0.3\textwidth]{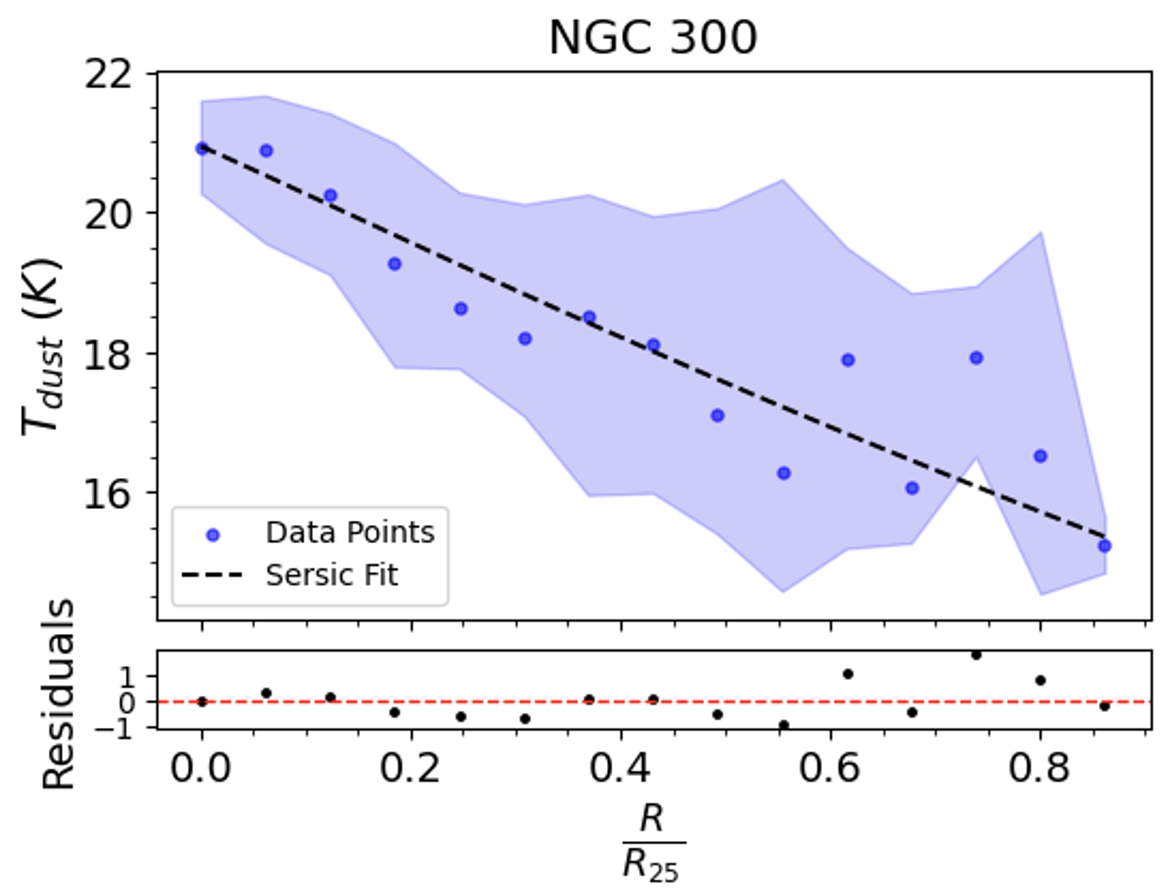} &
\includegraphics[width=0.3\textwidth]{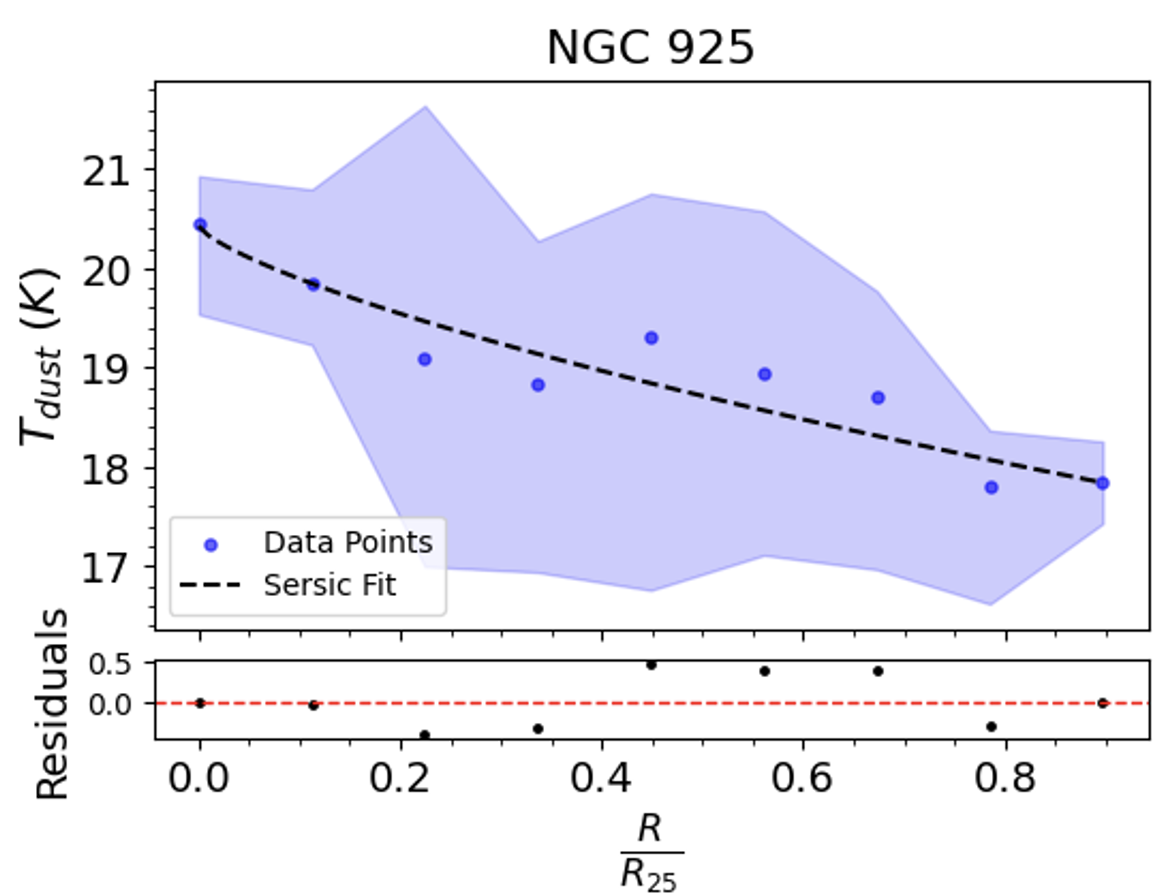} &
\includegraphics[width=0.3\textwidth]{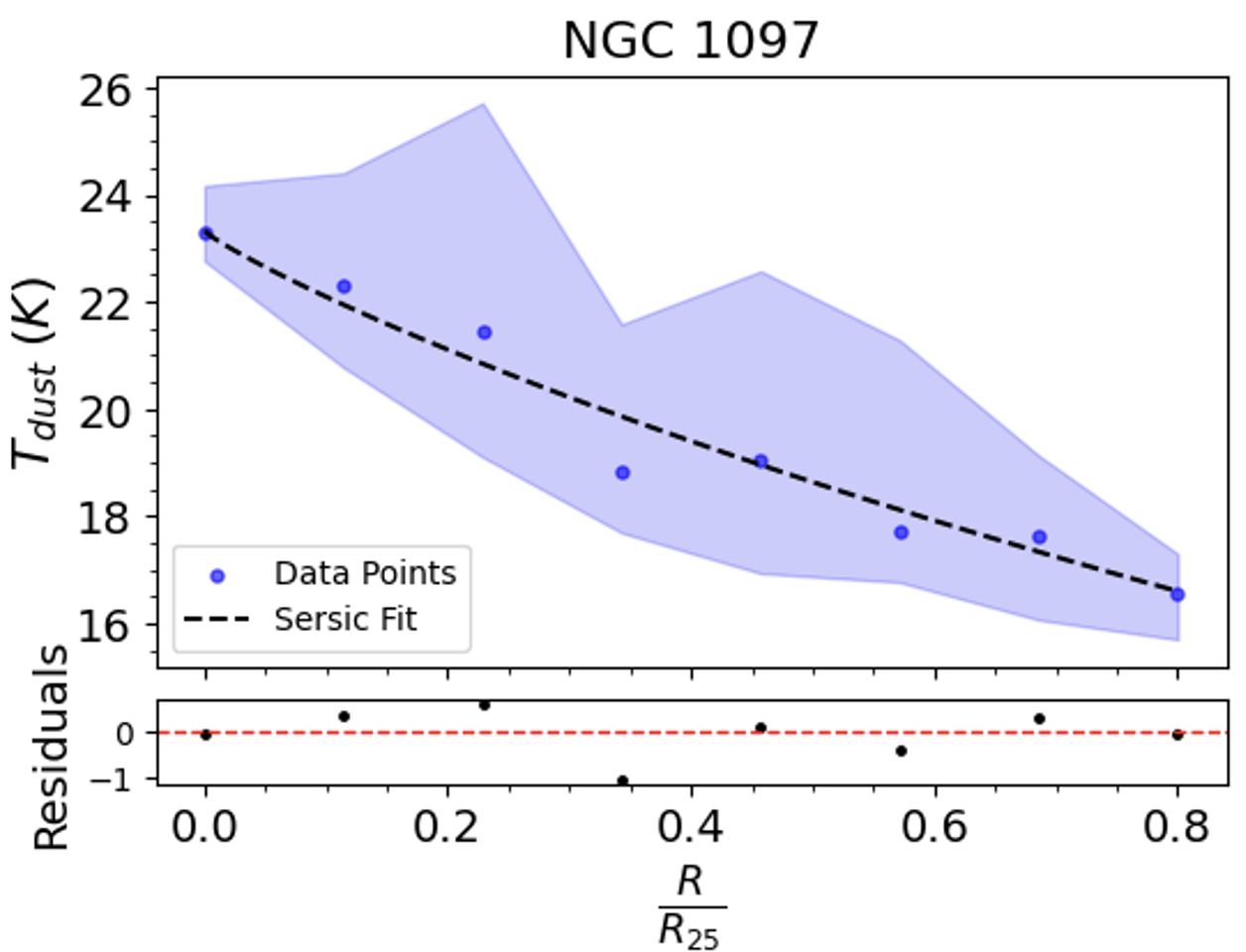}\\[1ex]
\includegraphics[width=0.3\textwidth]{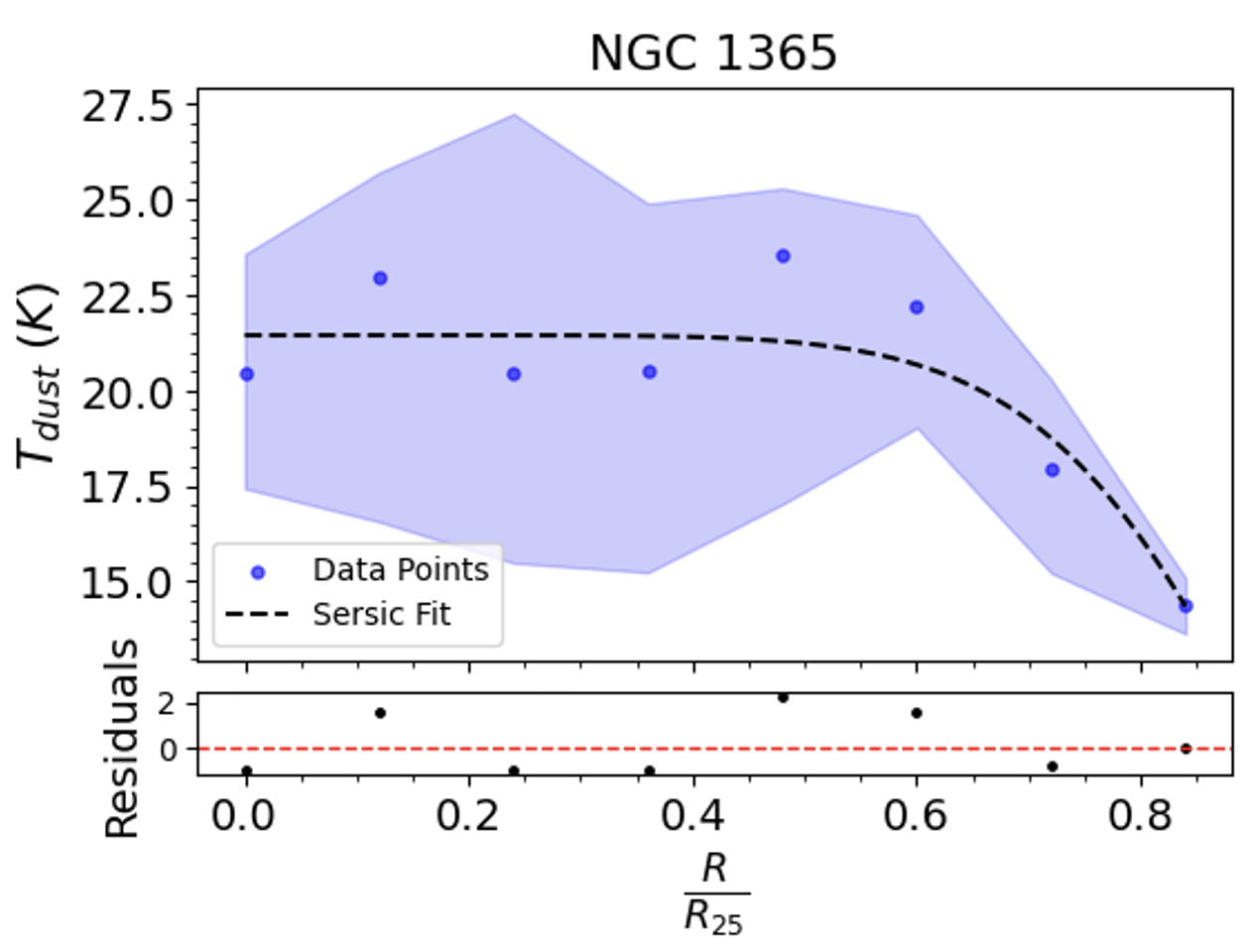} &
\includegraphics[width=0.3\textwidth]{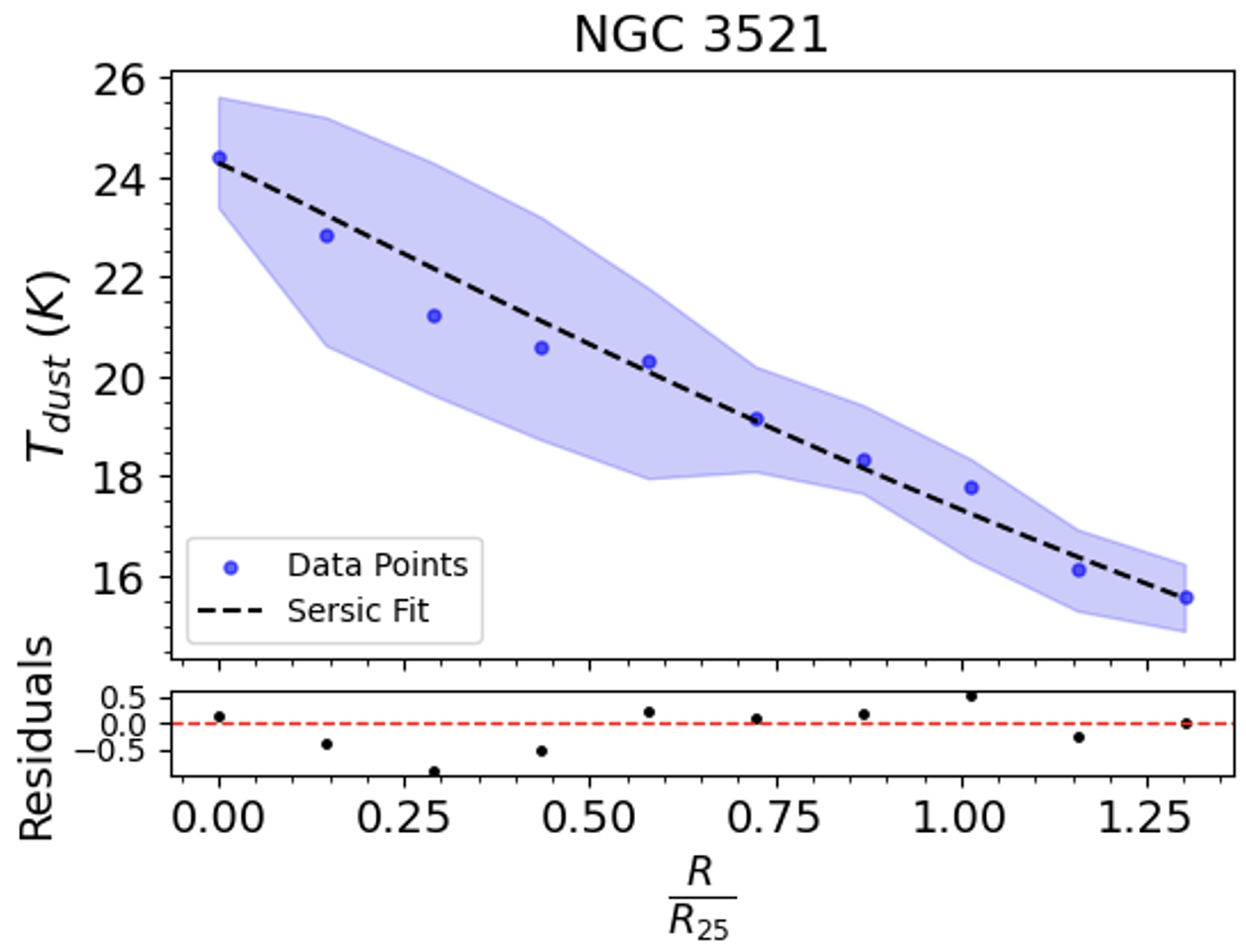} &
\includegraphics[width=0.3\textwidth]{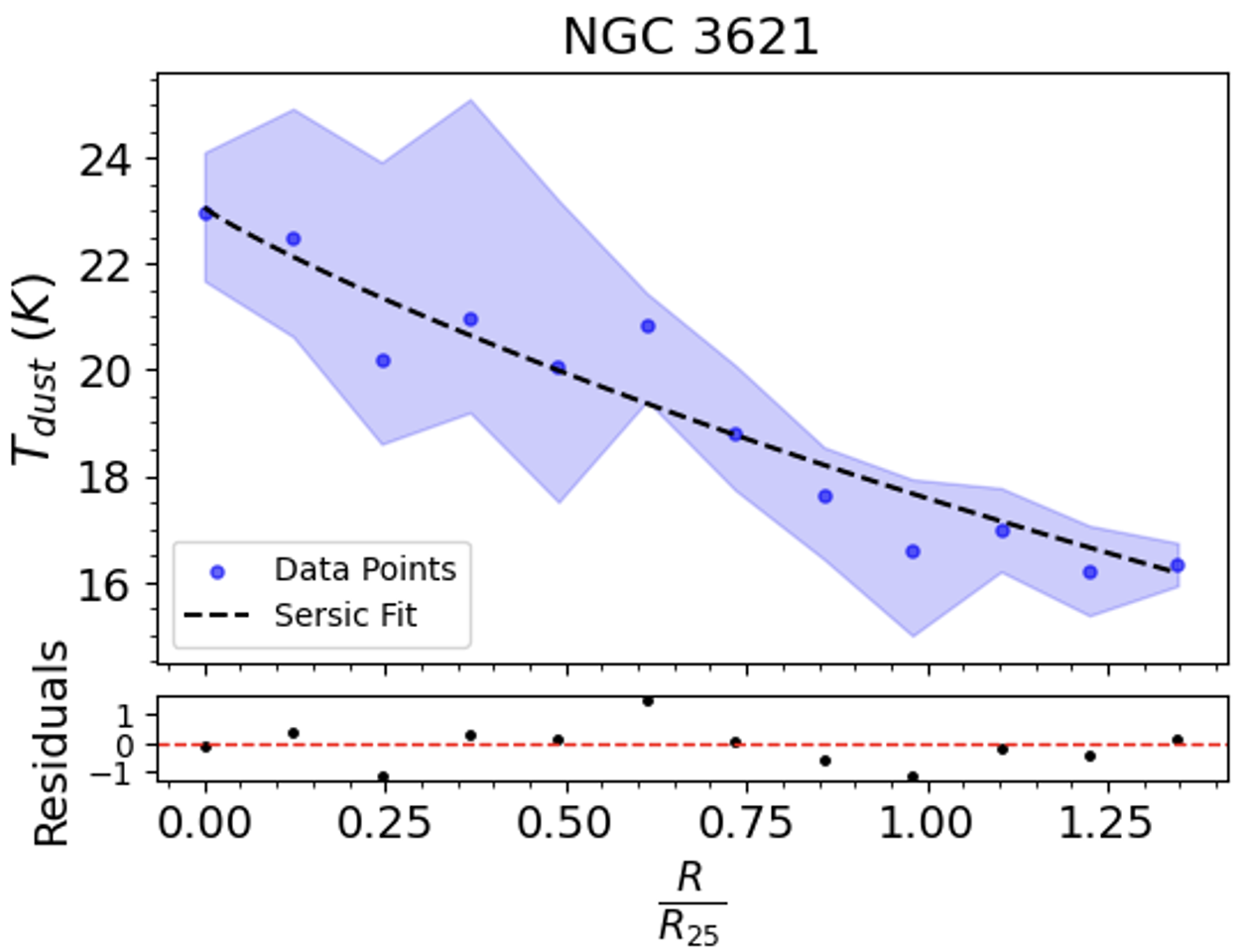}\\[1ex]
\includegraphics[width=0.3\textwidth]{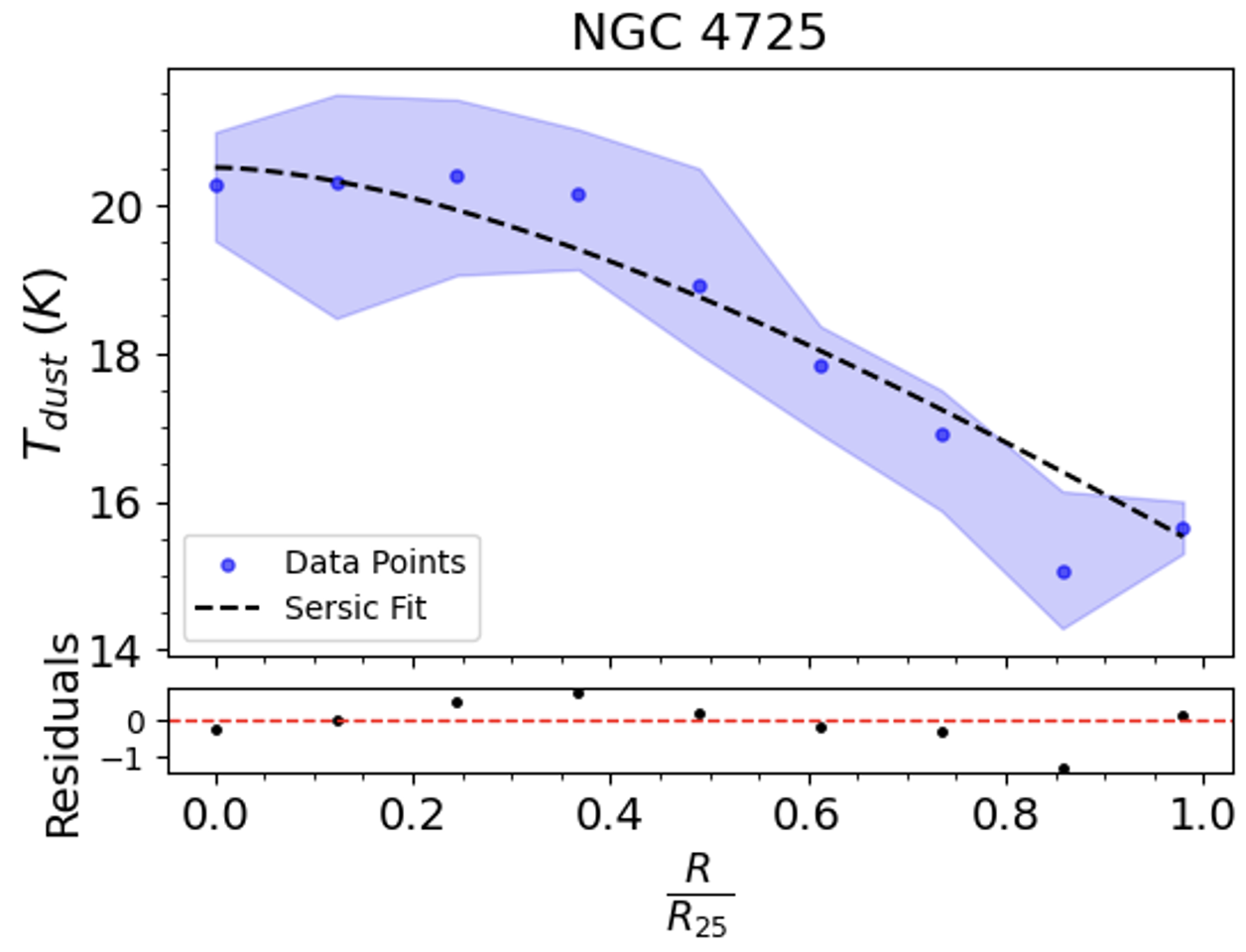}&
\includegraphics[width=0.3\textwidth]{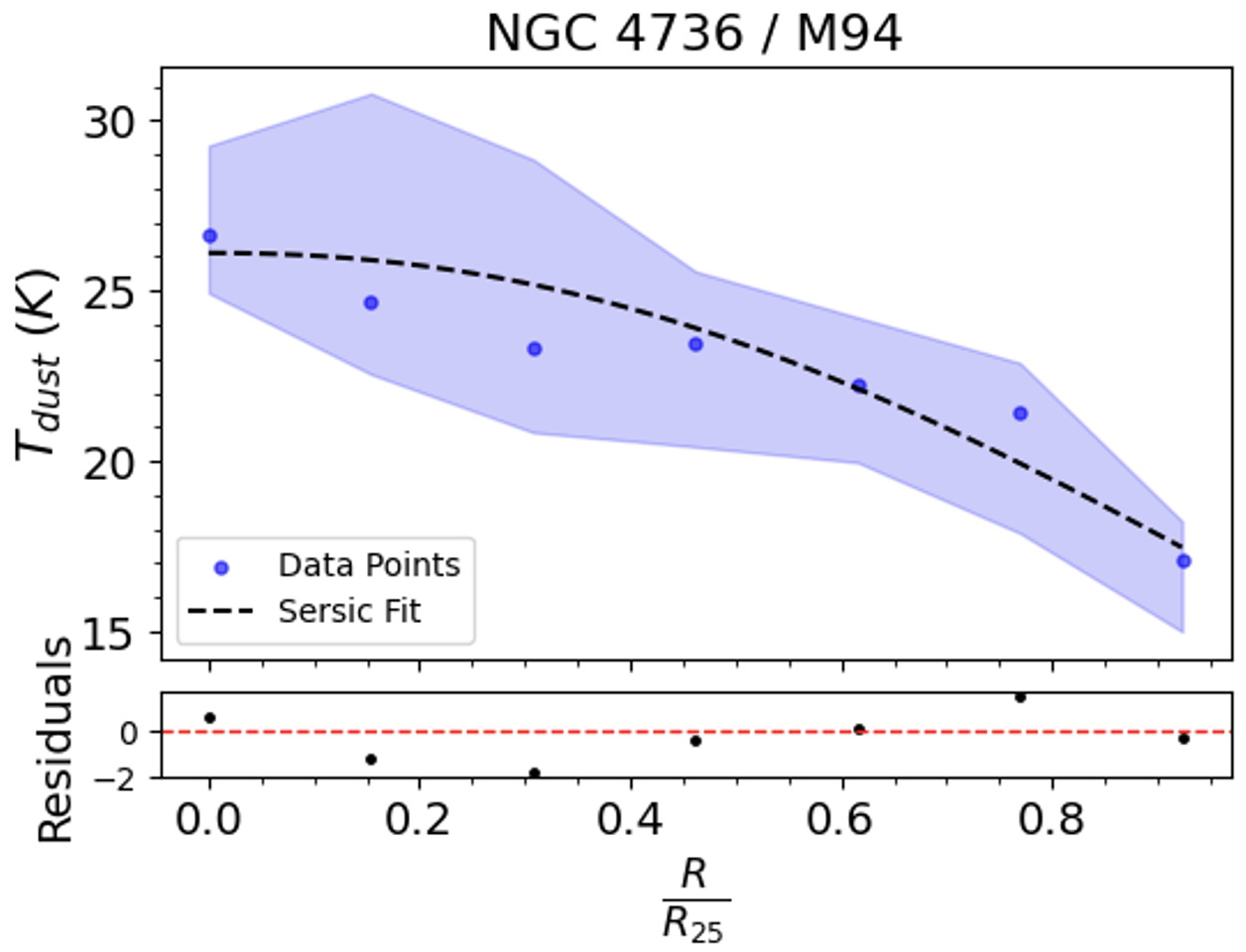} &
\includegraphics[width=0.3\textwidth]{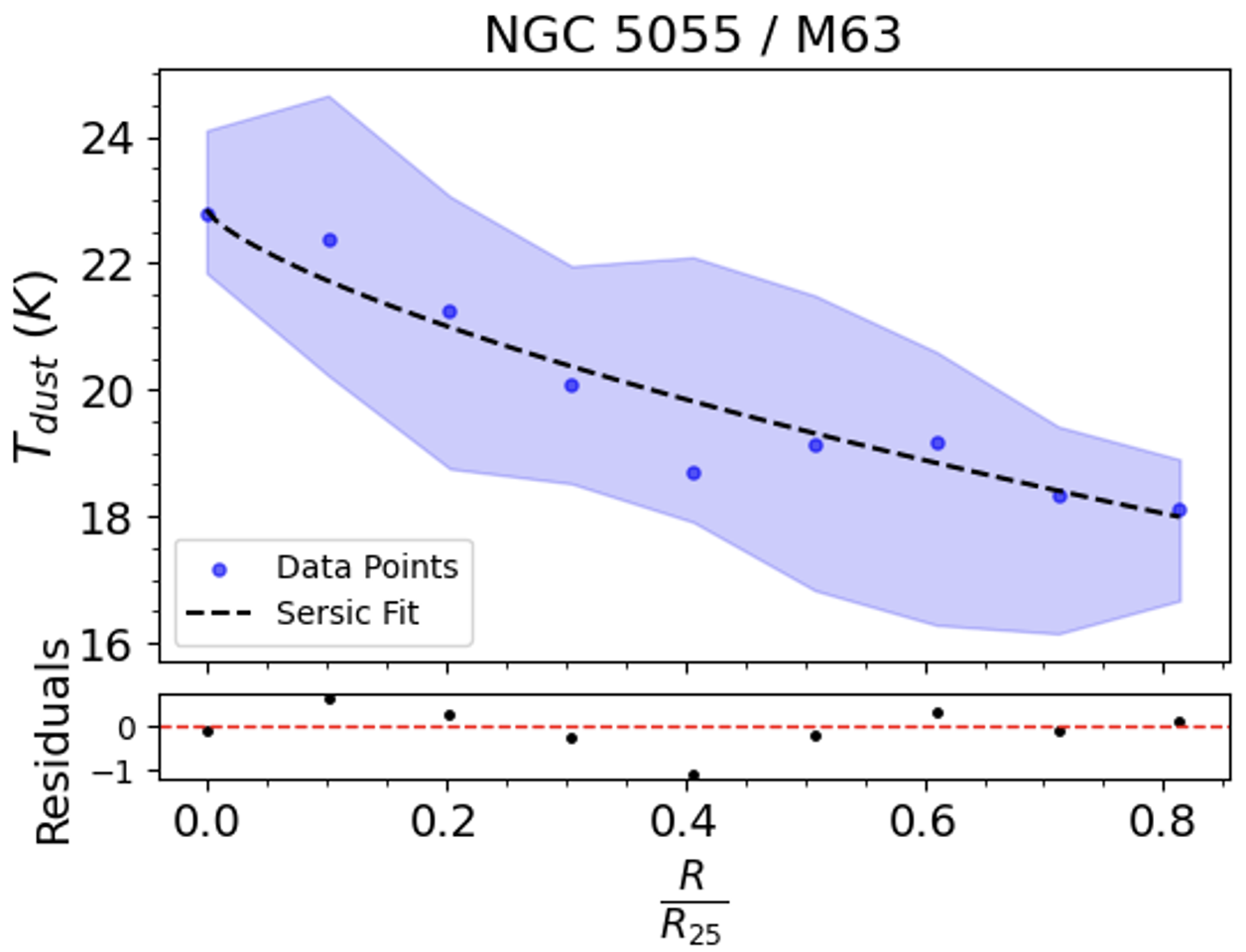} \\[1ex]
\includegraphics[width=0.3\textwidth]{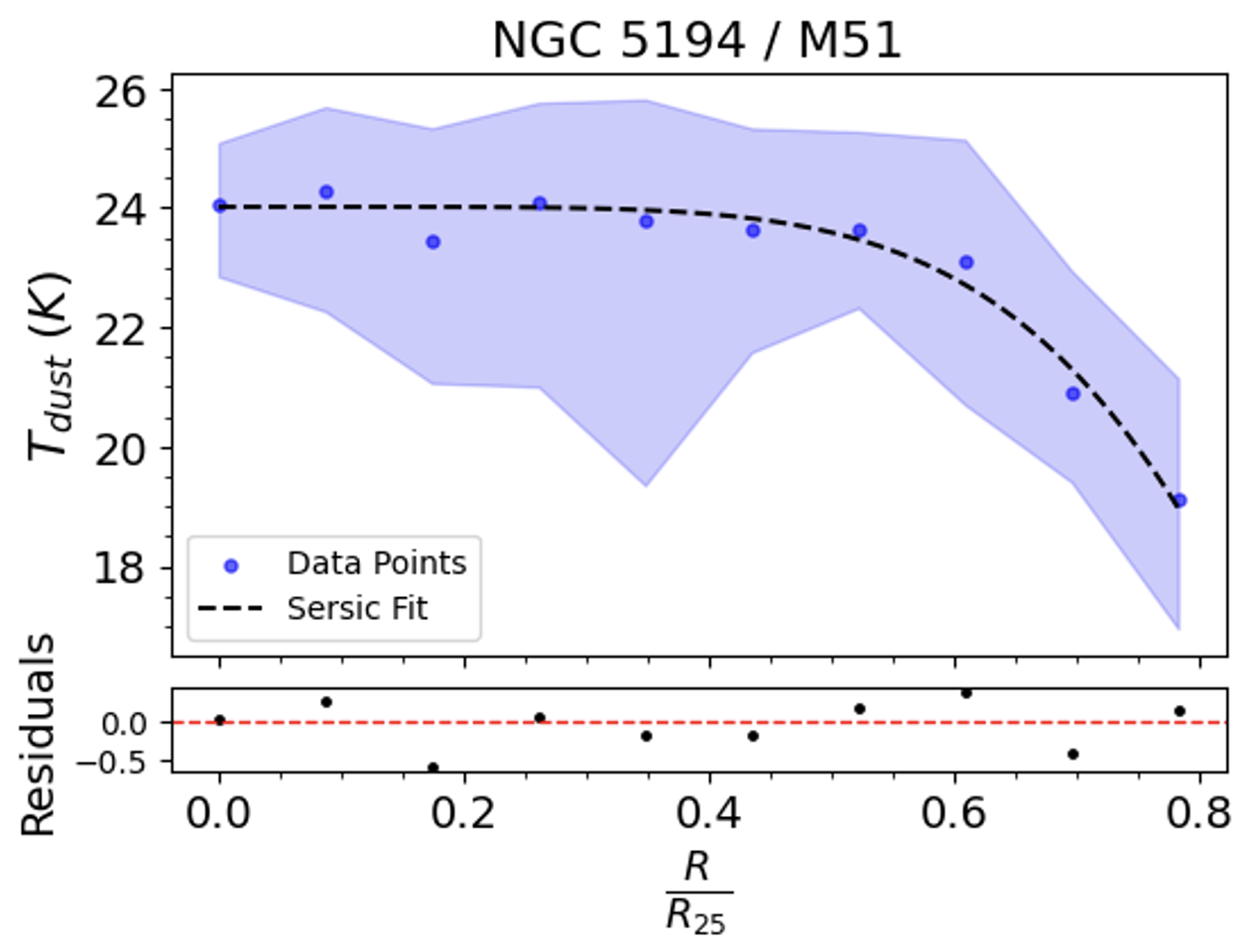}&
\includegraphics[width=0.3\textwidth]{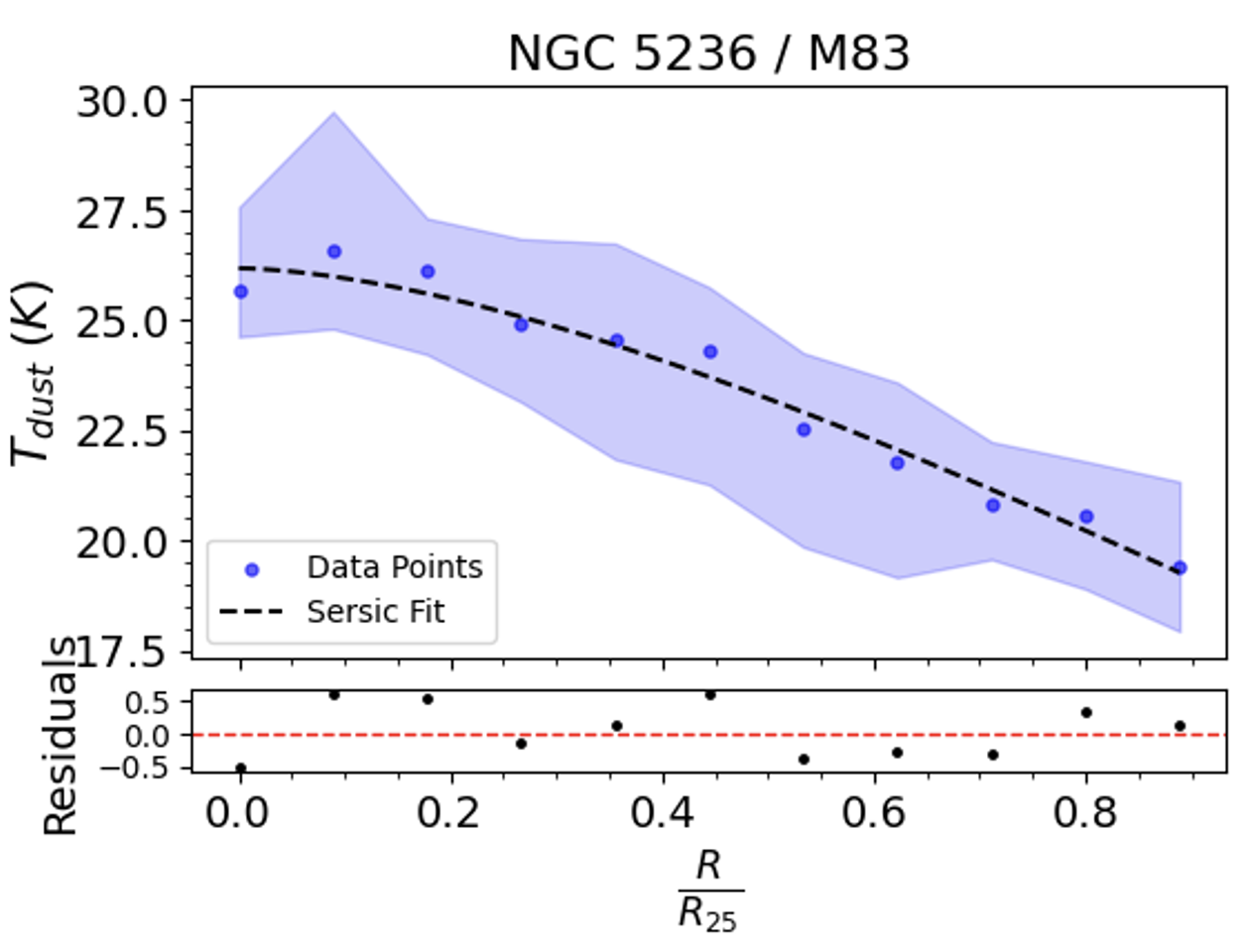} &
\includegraphics[width=0.3\textwidth]{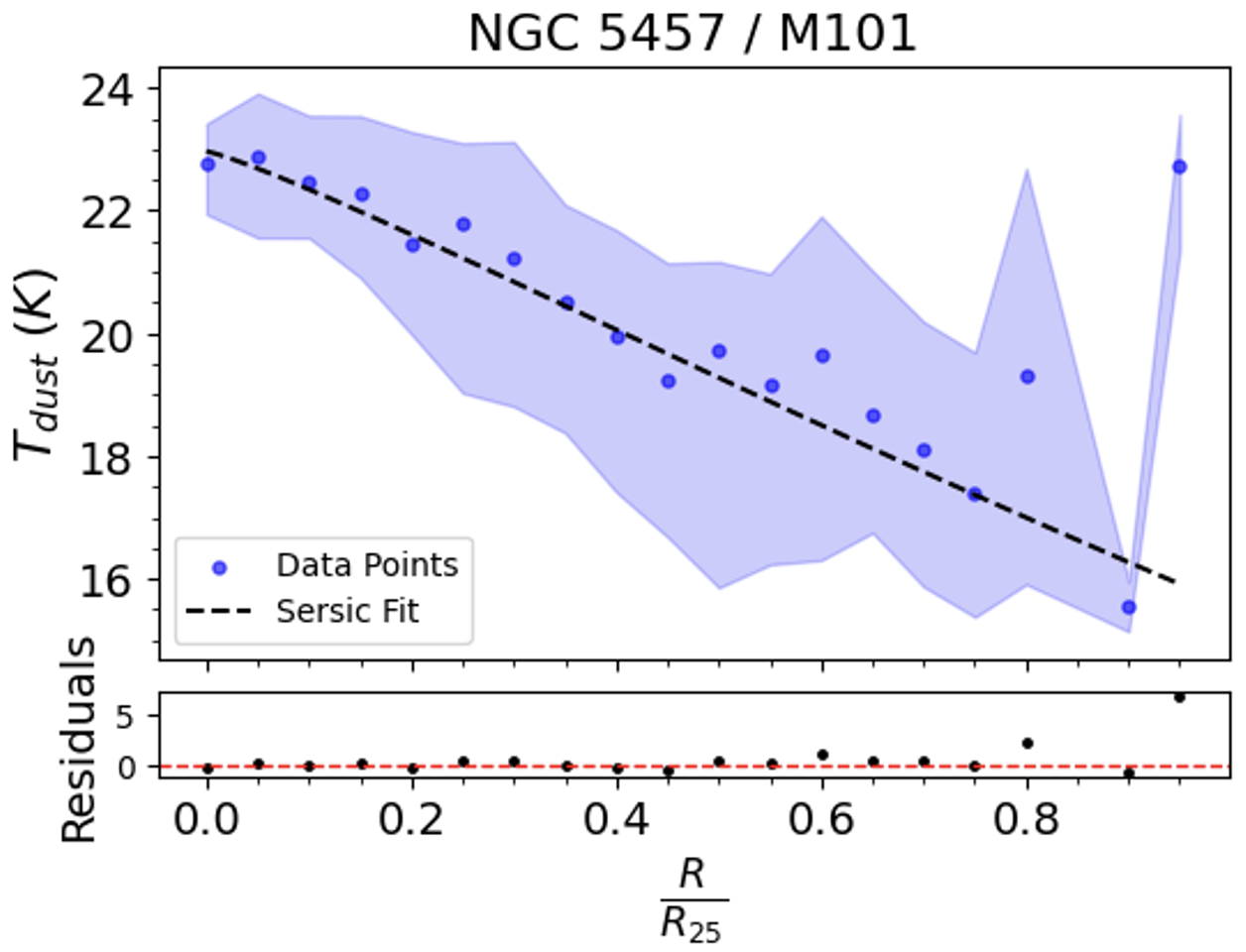} \\[1ex]
\includegraphics[width=0.3\textwidth]{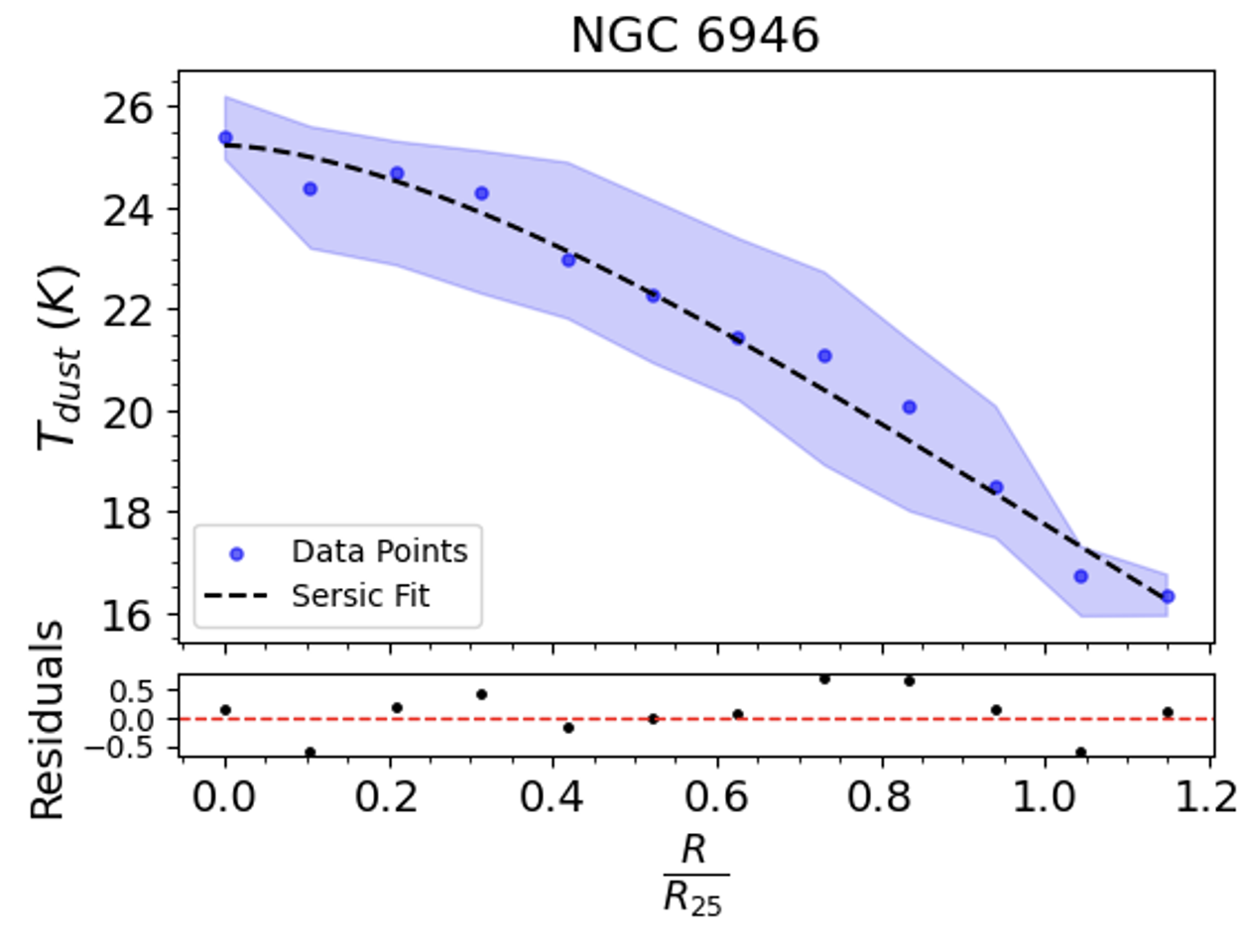}&
\includegraphics[width=0.3\textwidth]{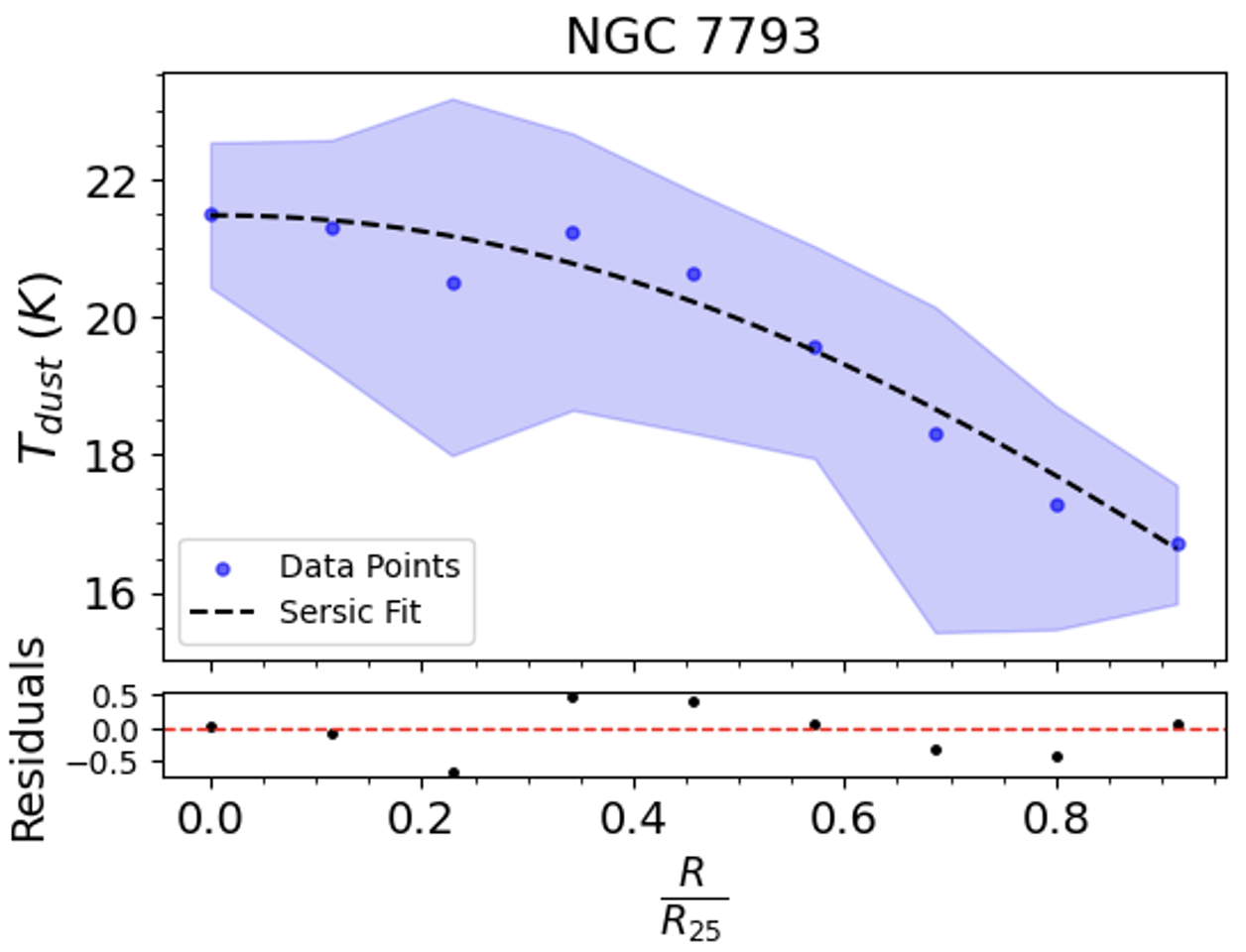} &
\end{tabular}
\caption{$T_{\rm dust}$ radial profiles as a function of the galaxy radius normalized by $R_{25}$ for our sample except for IC~342, NGC~628, NGC~2403, and NGC~3031, already displayed in Fig.~\ref{figradprof}.}
\label{appfigradprof}
\end{figure}

%\clearpage
\section{Images of Method 2}
\label{appendix: Method 2}
\begin{figure}[htbp]
\centering
\begin{tabular}{ccc}
\includegraphics[width=0.3\textwidth]{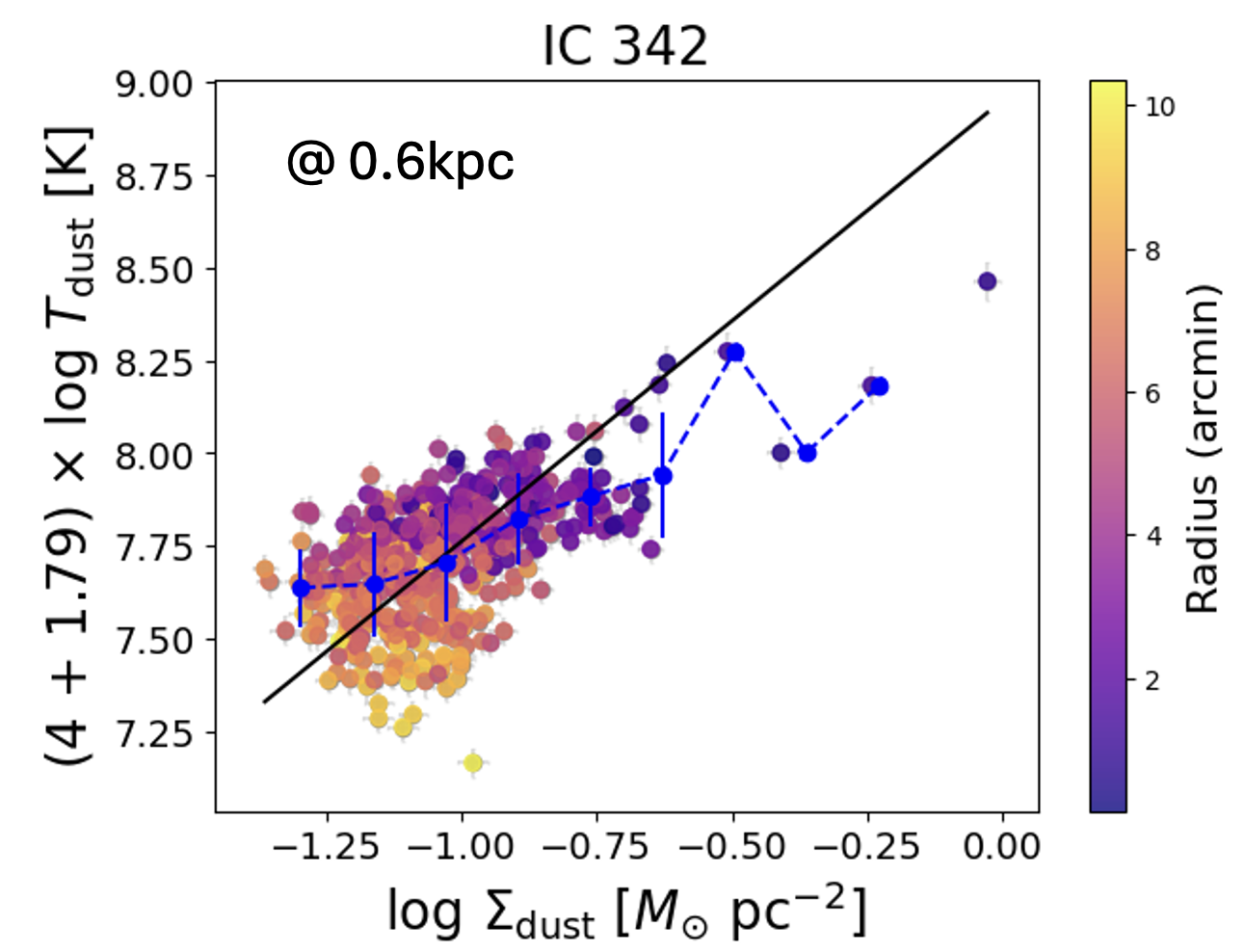} &
\includegraphics[width=0.3\textwidth]{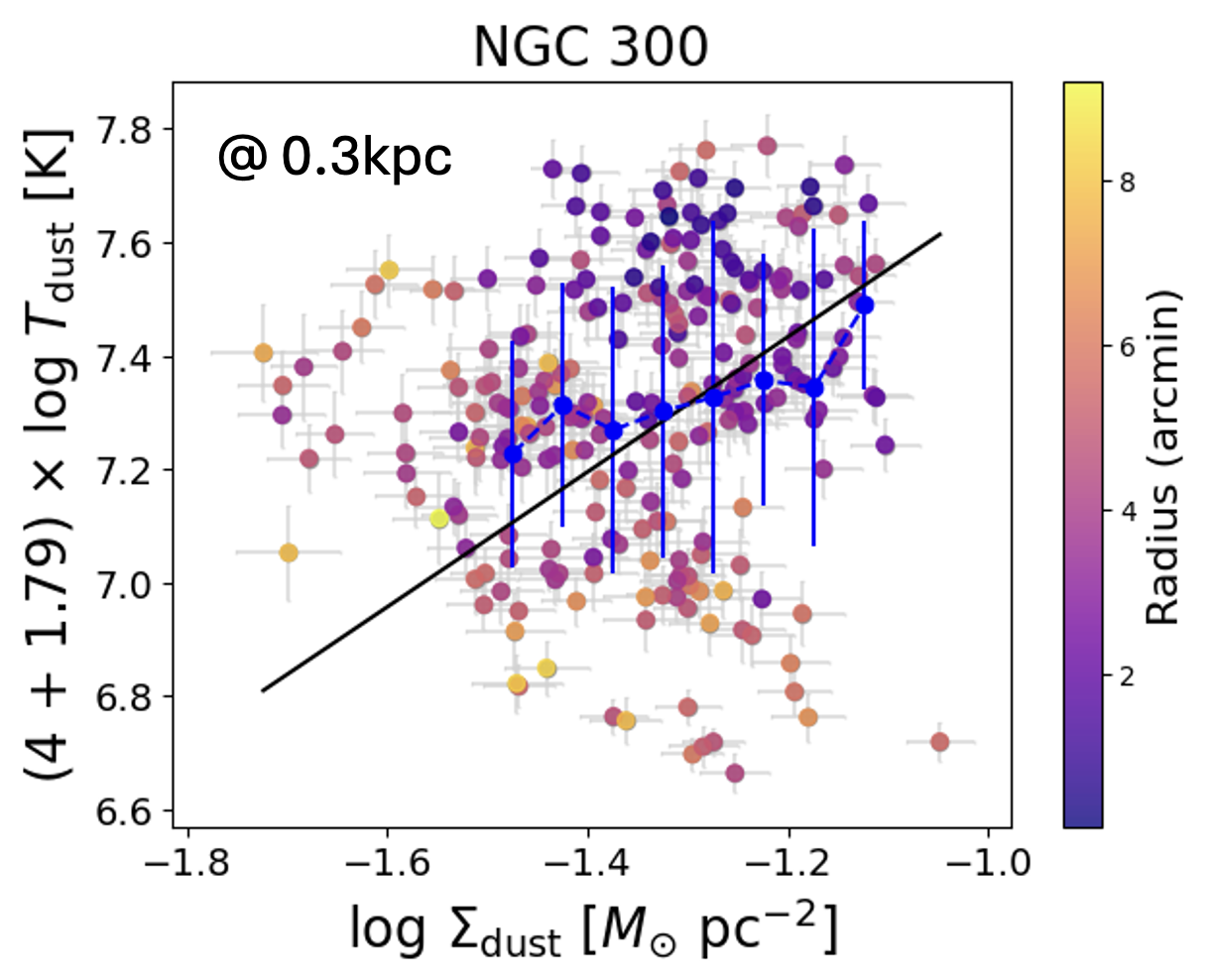} &
\includegraphics[width=0.3\textwidth]{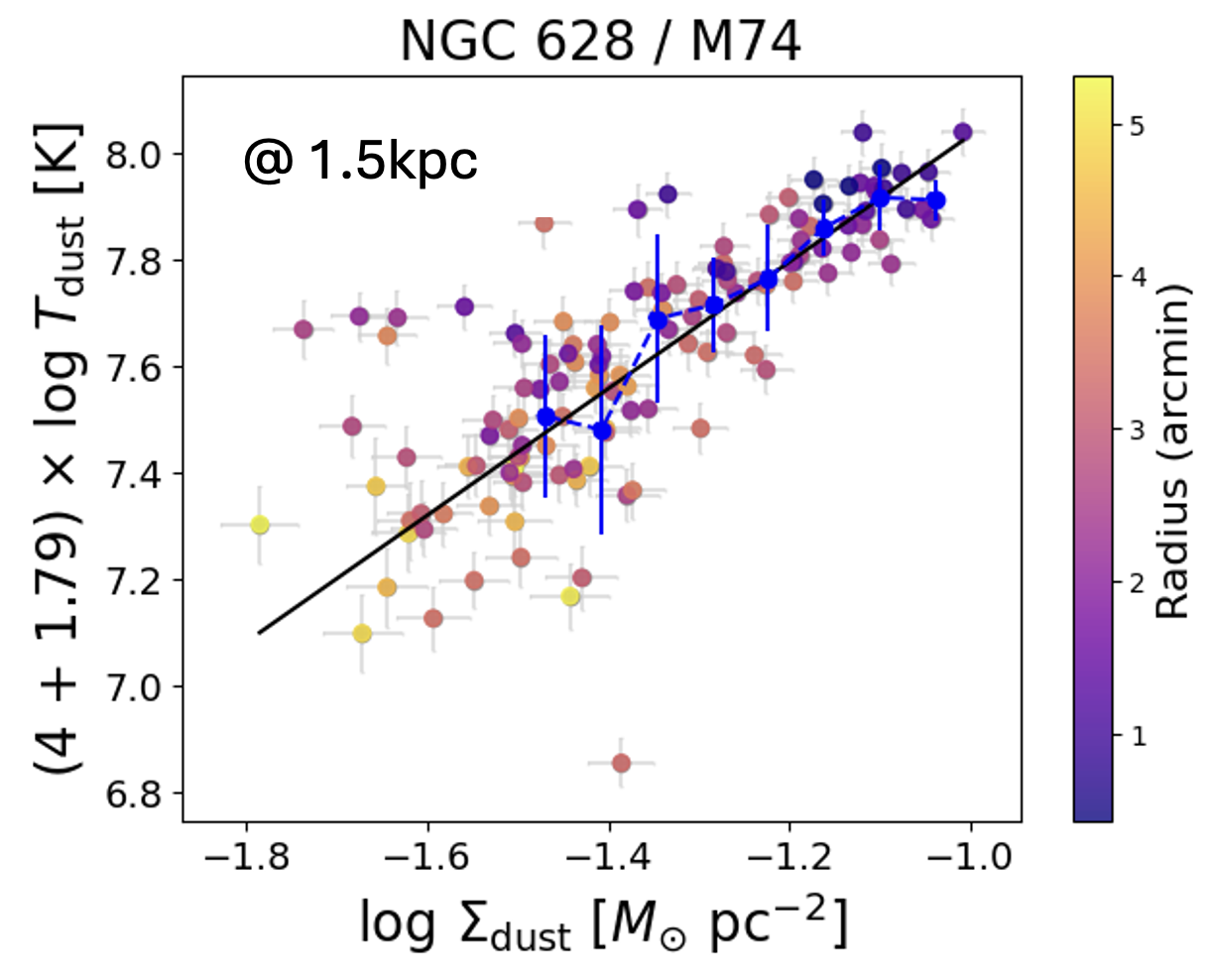}\\[1ex]
\includegraphics[width=0.3\textwidth]{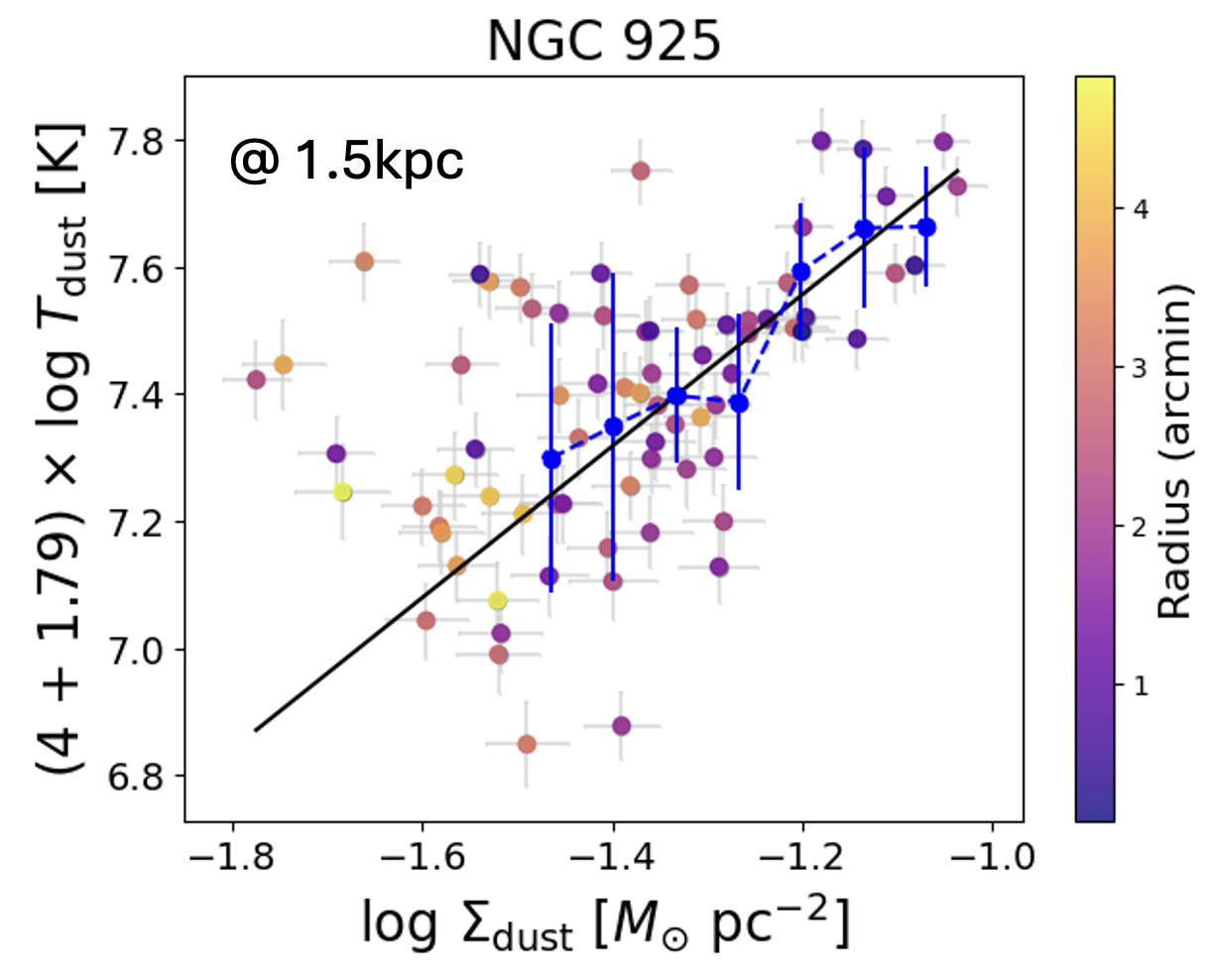} &
\includegraphics[width=0.3\textwidth]{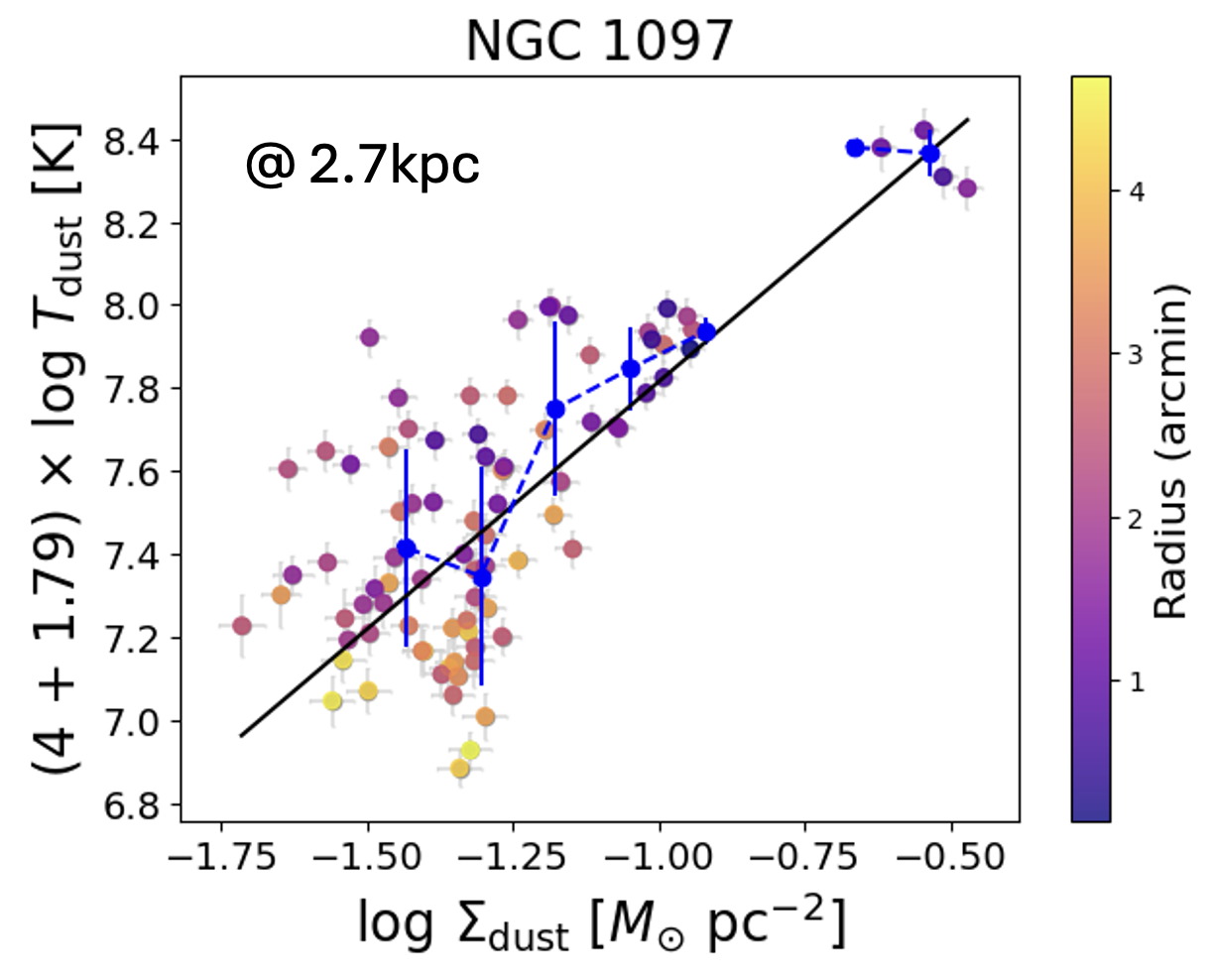} &
\includegraphics[width=0.3\textwidth]{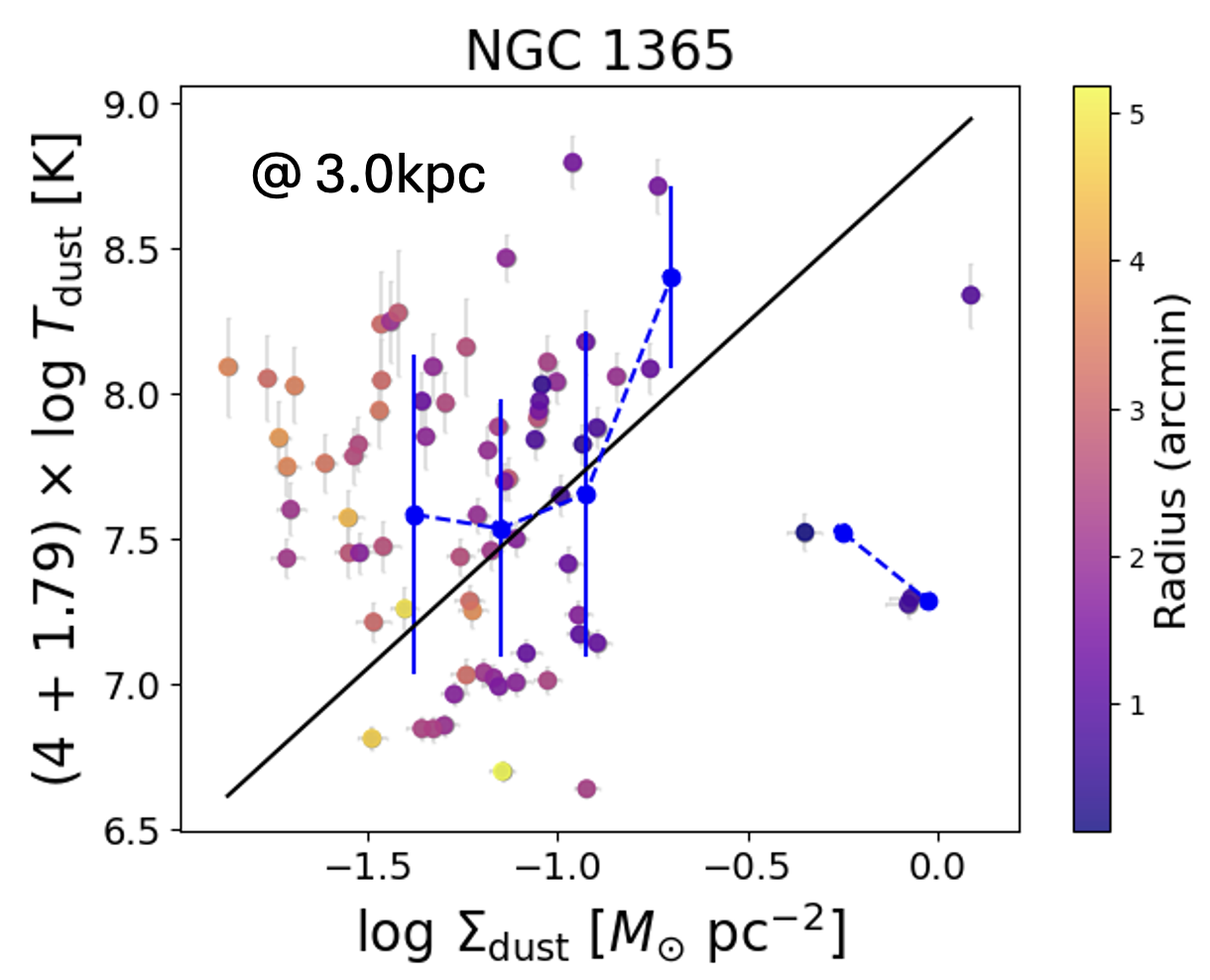}\\[1ex]
\includegraphics[width=0.3\textwidth]{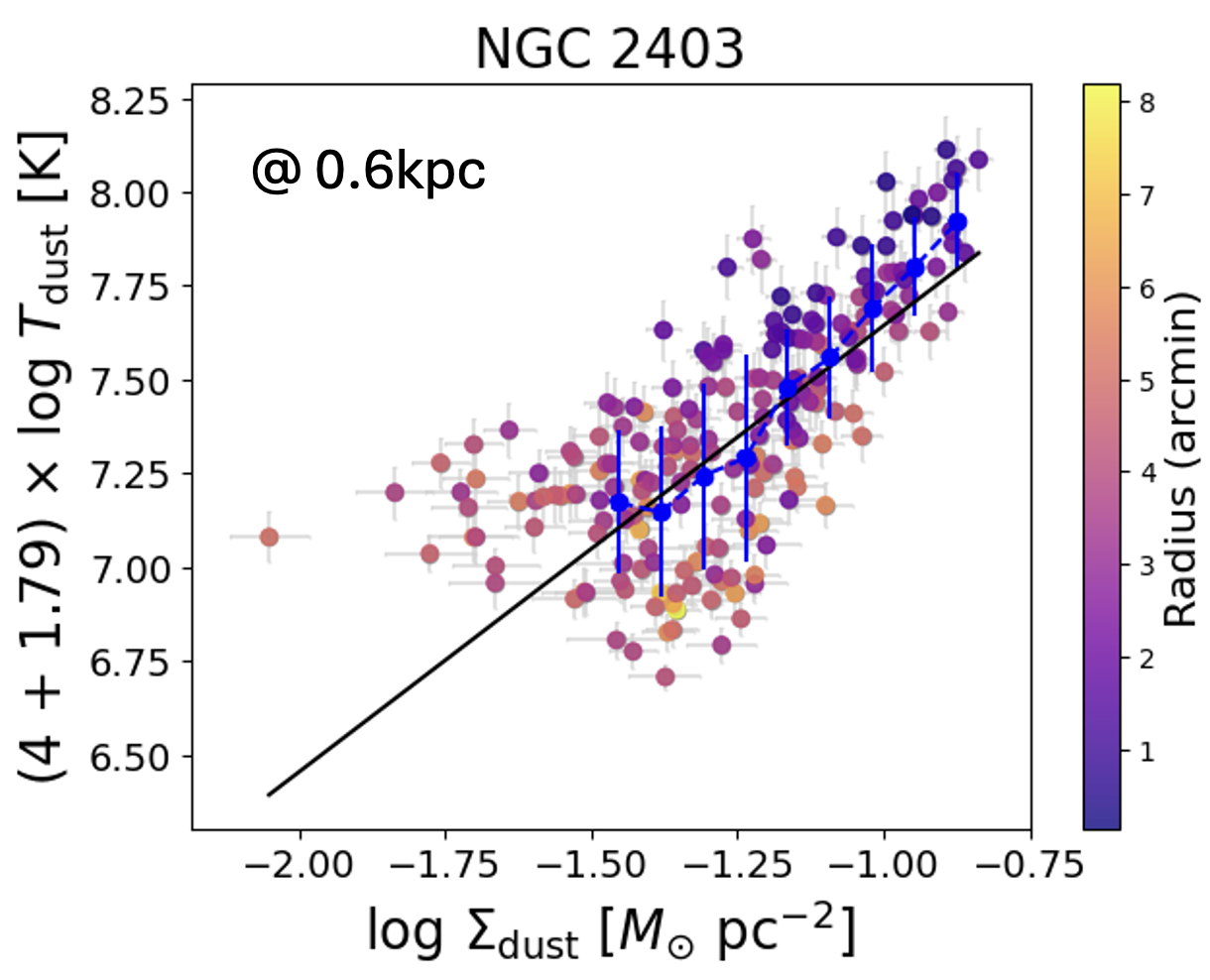}&
\includegraphics[width=0.3\textwidth]{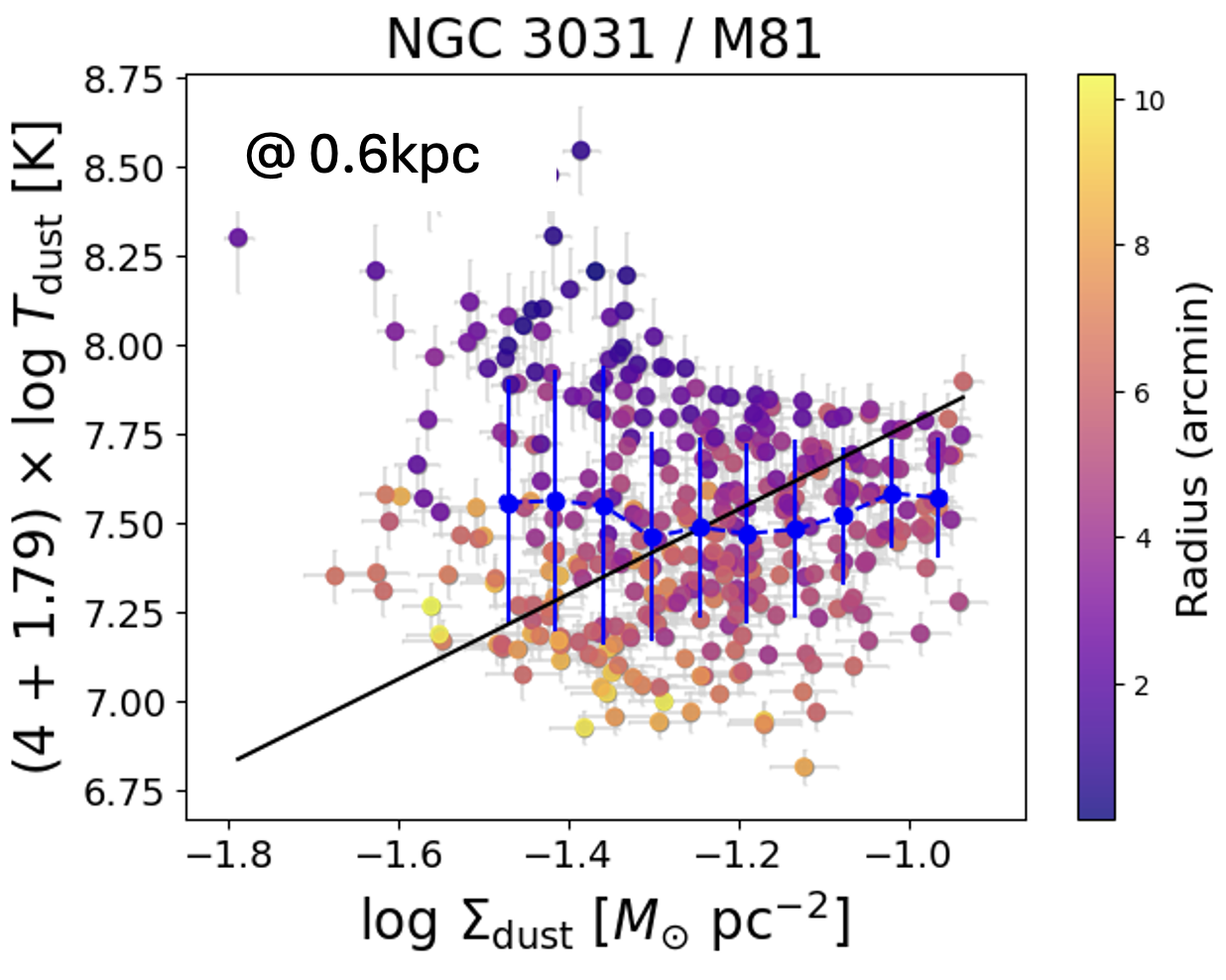} &
\includegraphics[width=0.3\textwidth]{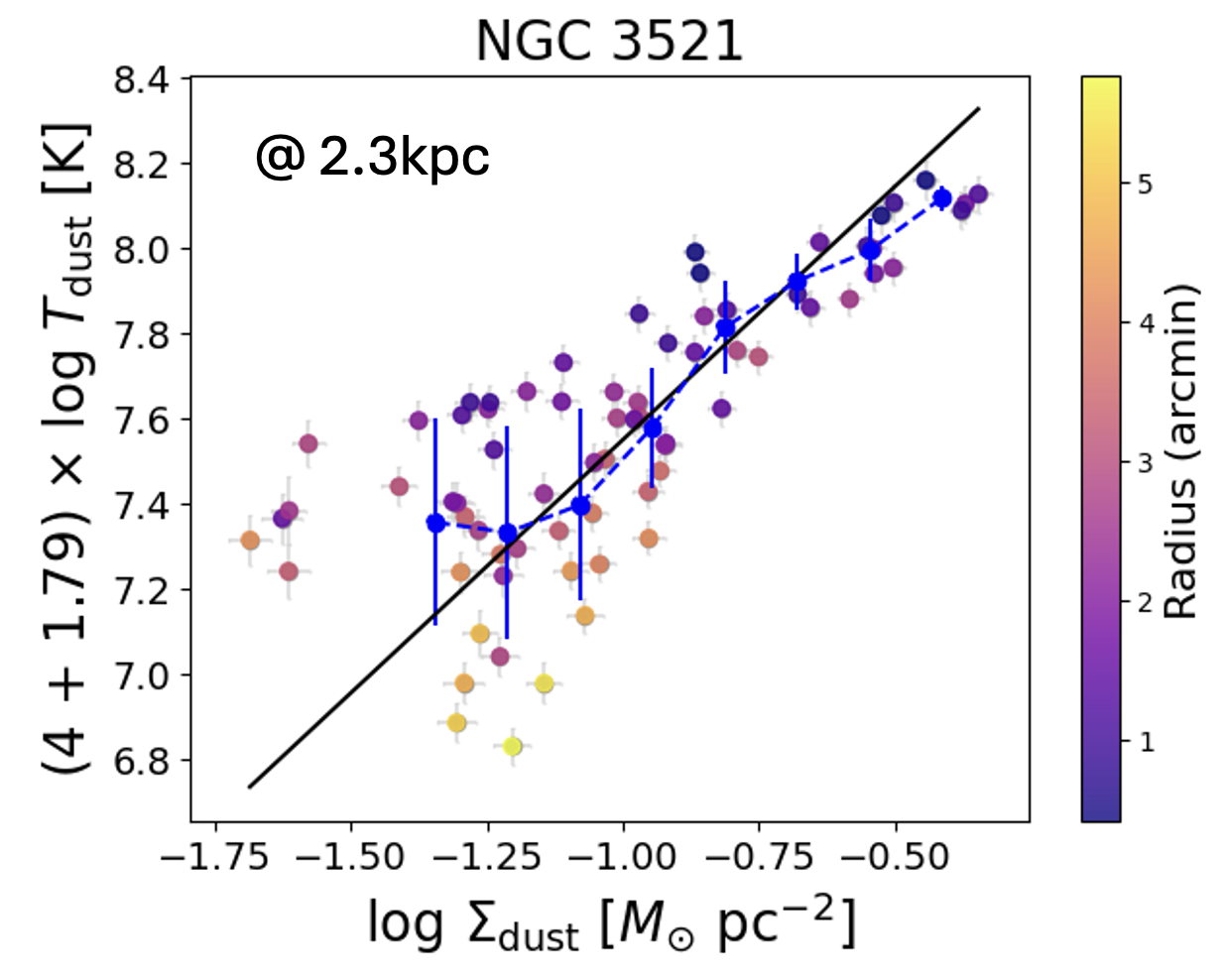} \\[1ex]
\includegraphics[width=0.3\textwidth]{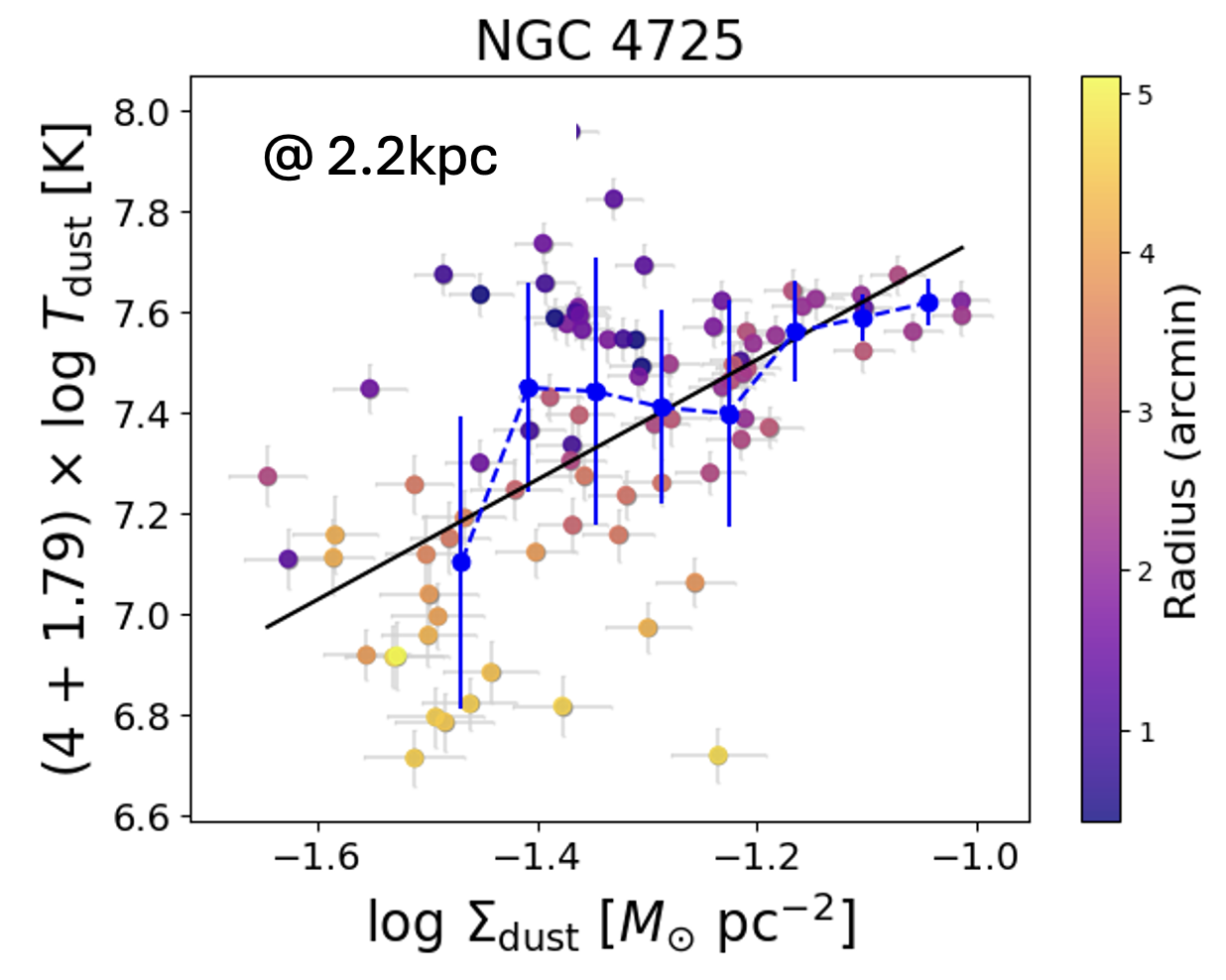}&
\includegraphics[width=0.3\textwidth]{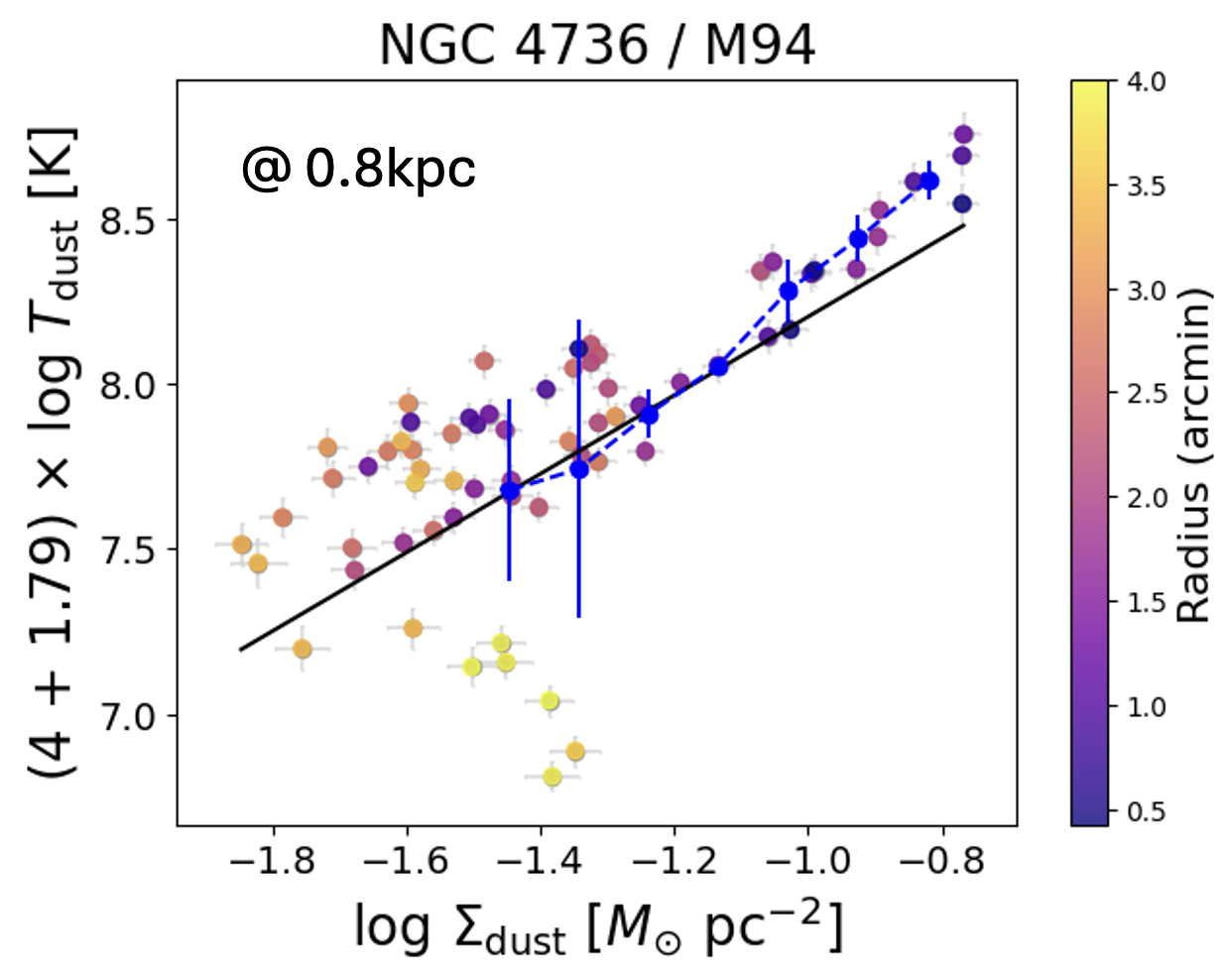} &
\includegraphics[width=0.3\textwidth]{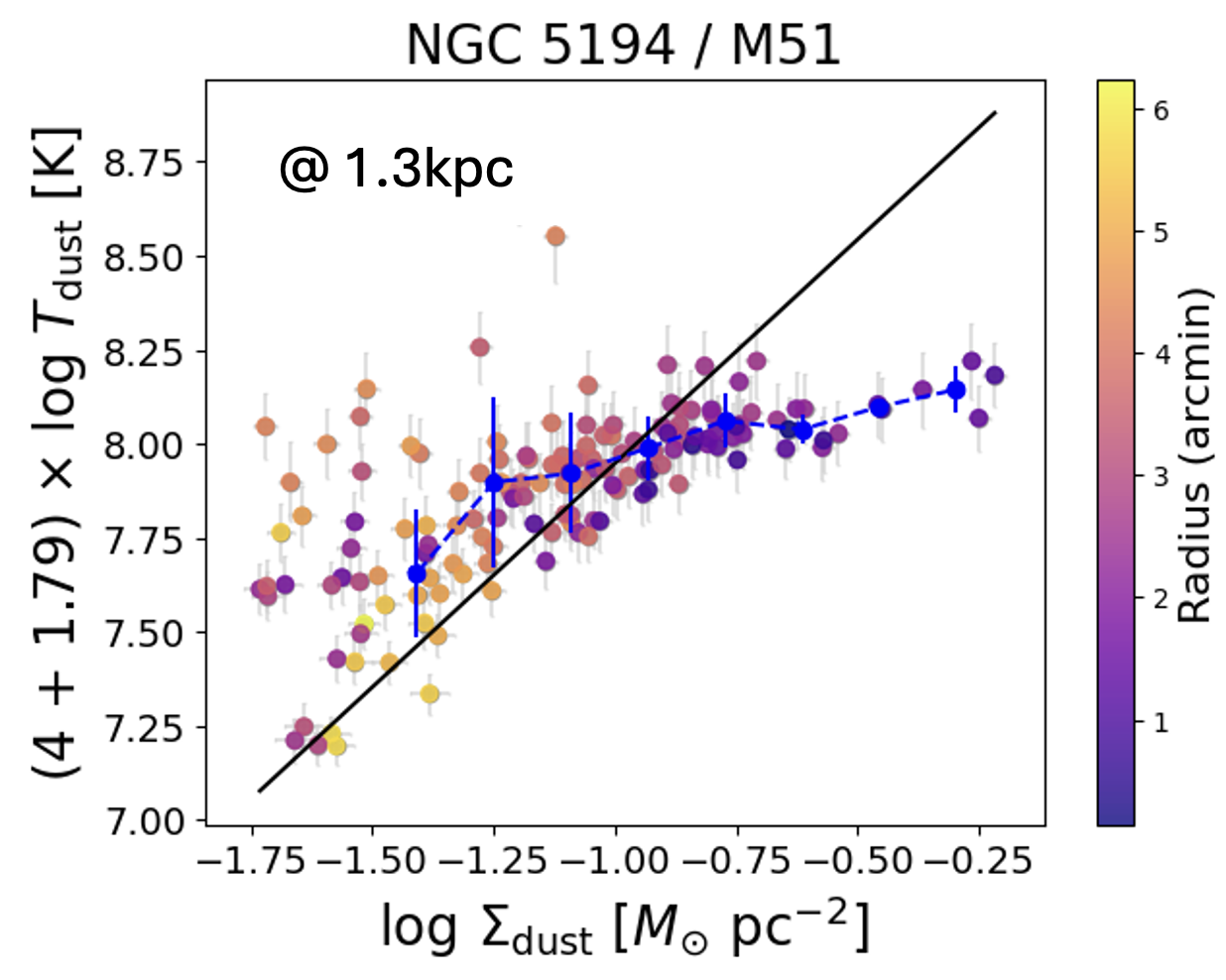} \\[1ex]
\end{tabular}
\caption{The pixel-by-pixel relation between $T_{\rm dust}$ and log\(\Sigma_{dust}\) according to Eq.(\ref{eq:temp_dust_final}) for our sample except for NGC~3621 and NGC~5055, already displayed in Fig.~\ref{figdust}.}
\label{appT-dustmass}
\end{figure}
\begin{figure}[htbp]
\ContinuedFloat
\centering
\begin{tabular}{ccc}
\includegraphics[width=0.3\textwidth]{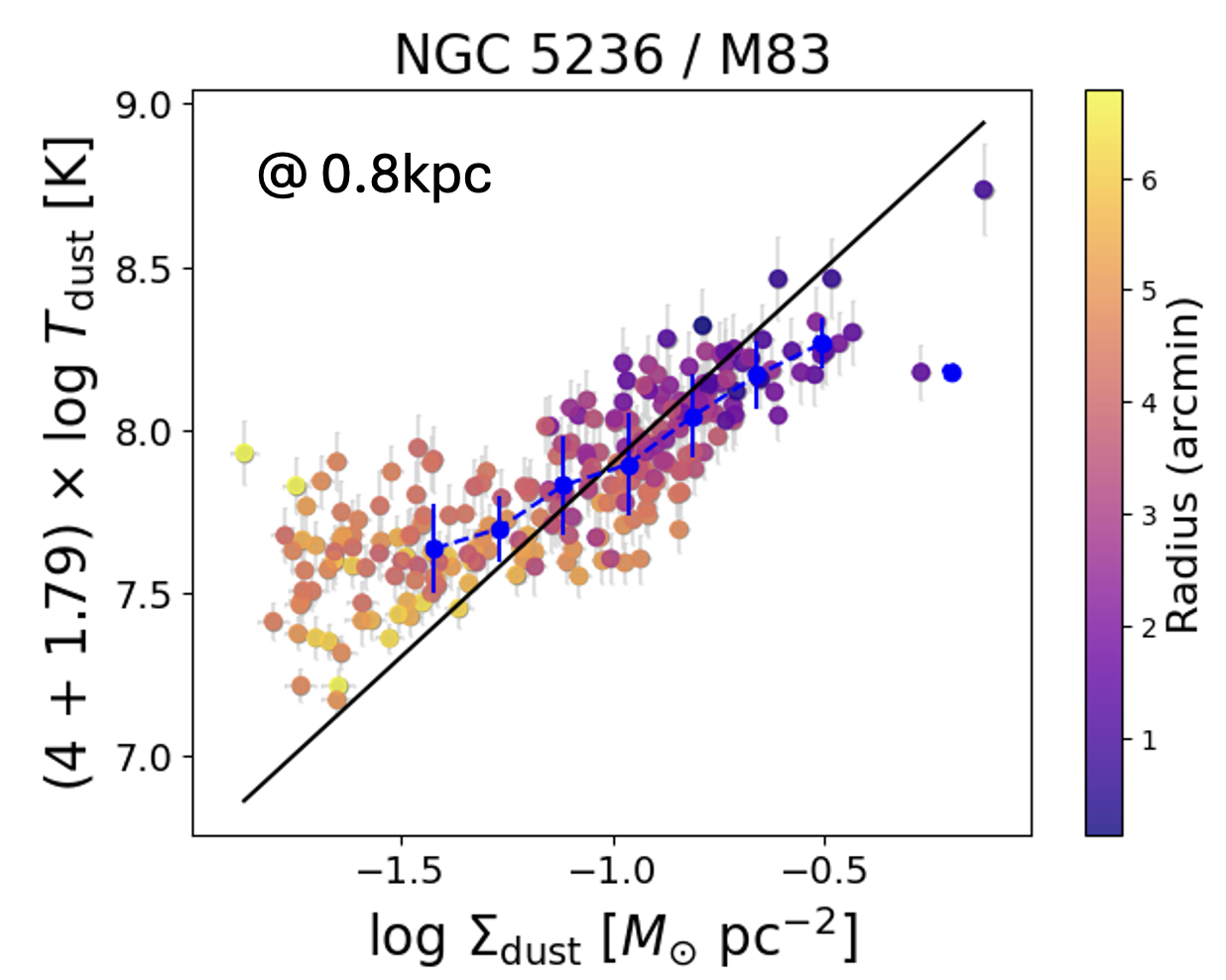}&
\includegraphics[width=0.3\textwidth]{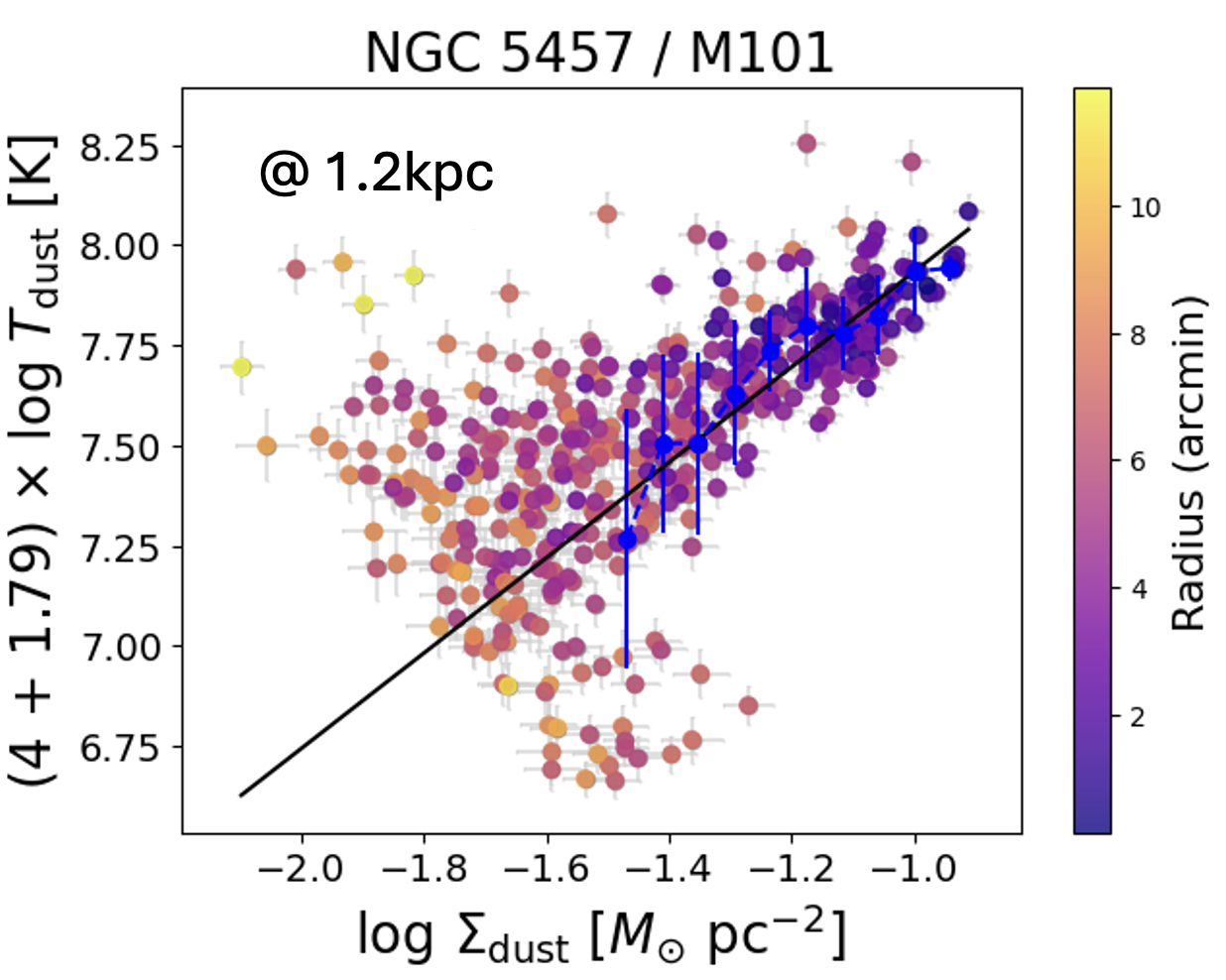} &
\includegraphics[width=0.3\textwidth]{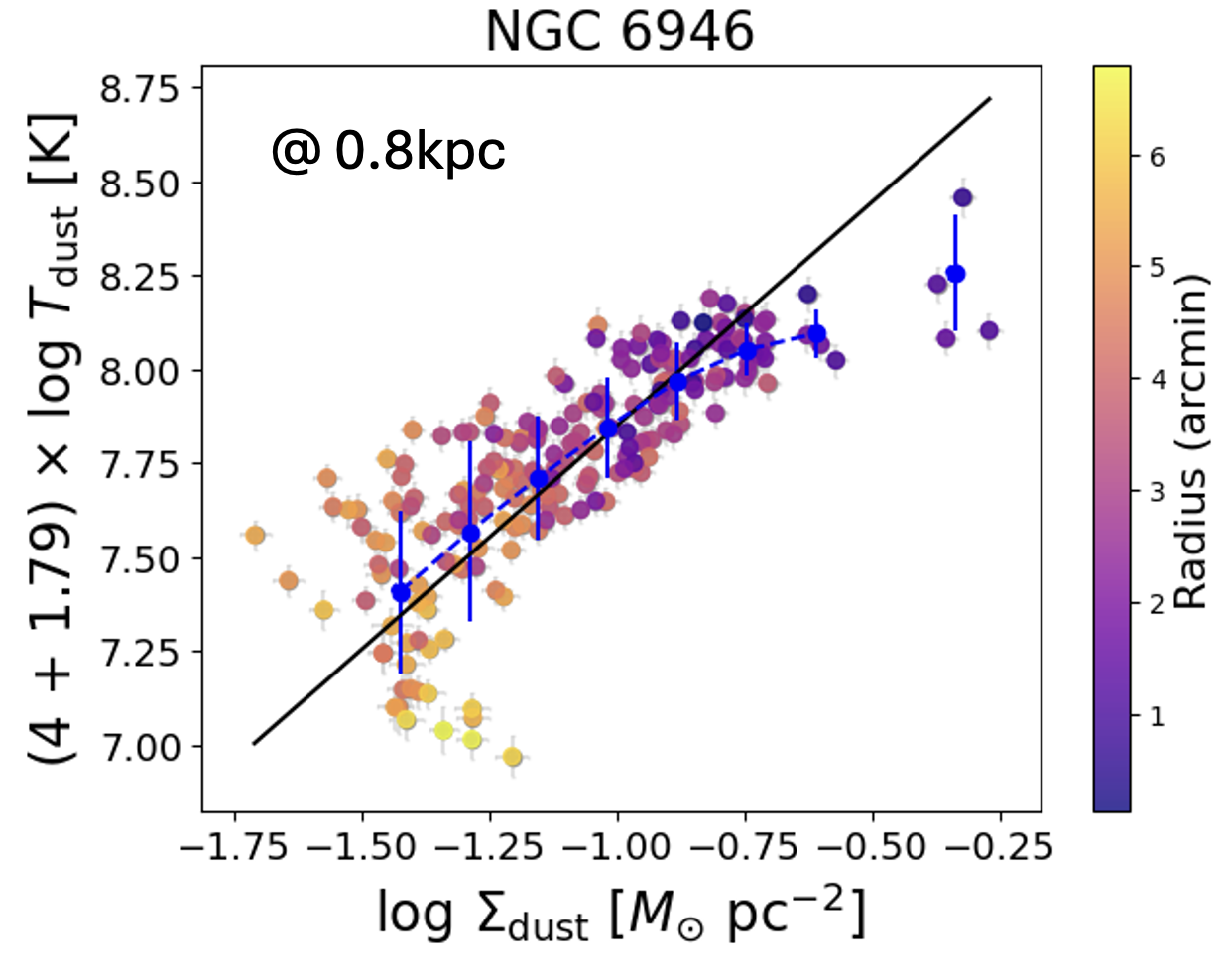}\\[1ex]
&
\includegraphics[width=0.3\textwidth]{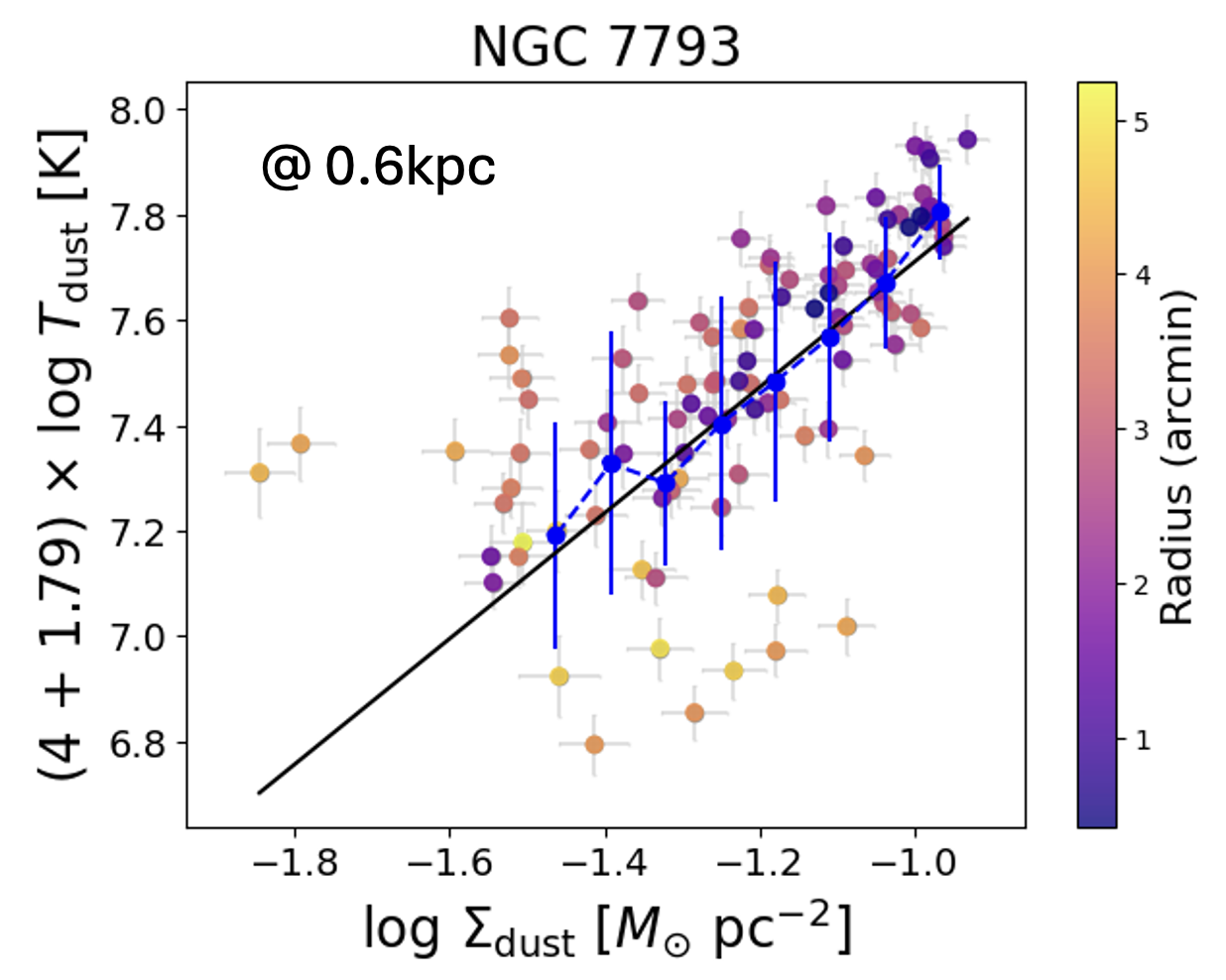} &
\end{tabular}
\caption{\textit{continued.}}
\end{figure}
\end{document}